\documentclass[aps,twocolumn]{revtex4}
\usepackage{graphicx}    \usepackage{rotating}    \usepackage{dcolumn}
\usepackage{epsfig} 

\begin{document}  

\title{Optimal network topologies for information transmission in active networks}

\author{M.  S.  Baptista$^{1}$, J. X. de Carvalho$^{1}$,
M. S. Hussein$^{1,2}$}
\affiliation{$^1$Max-Planck-Institut
f\"ur Physik komplexer Systeme, N\"othnitzerstr. 38, D-01187 Dresden,
Deutschland}
\affiliation{$^2$Institute of Physics, University of S\~ao Paulo\\
Rua do Mat\~ao, Travessa R, 187, 05508-090, S\~ao Paulo - Brasil}
\date{\today}

\begin{abstract}
  This work clarifies the relation between network circuit (topology)
  and behavior (information transmission and synchronization) in active
  networks, e.g. neural networks. As an application, we show how to determine
  a network topology that is optimal for information transmission.  By
  optimal, we mean that the network is able to transmit a large amount of
  information, it possesses a large number of communication channels, and it
  is robust under large variations of the network coupling configuration. This
  theoretical approach is general and does not depend on the particular dynamic
  of the elements forming the network, since the network topology can be
  determined by finding a Laplacian matrix (the matrix that describes the
  connections and the coupling strengths among the elements) whose eigenvalues
  satisfy some special conditions. To illustrate our ideas and theoretical
  approaches, we use neural networks of electrically connected chaotic
  Hindmarsh-Rose neurons.
\end{abstract}

\maketitle 


\section{Author Summary}

The relation between neural circuits and behavior is a fundamental matter in
neuroscience. In this work, we present a theoretical approach that has the
potential to unravel such a relationship in terms of network topology,
information, and synchronization, in active networks, networks formed by
elements that are dynamical systems (such as neurons, chaotic or periodic
oscillators). As a direct application of our approaches, we show how one can
construct optimal neural networks that not only transmit large amounts of
information from one element to another in the network, but also are robust
under alterations in the coupling configuration. We also show that the
relation between synchronization and information is rather subtle. Neural
networks whose configurations allow the transmission of large amounts of
information might have at least two unstable modes of oscillation that are out
of synchrony, while all the others are synchronous. Depending on the kind
of measurement being done, one can arrive at contradicting statements
concerning the relation between information and synchronization.  We
illustrate our theoretical approaches by using neural networks of
electrically connected chaotic Hindmarsh-Rose neurons
\cite{hindmarsh}. These results have a tremendous impact in the understanding
of information transmission in brain-like networks, as well as, in the
mammalian brain.  They also shed light on a better understanding of the
neural code, the rules under which neurons encode and transmit information
about external stimuli.

\section{Blurb}
This work shows how to relate in an active network the rate of information that
can be transmitted from one point to another, regarded as mutual information
rate (MIR), the synchronization level among elements, and the connecting
topology of the network.

\section{Introduction}

Given an arbitrary time dependent stimulus that externally excites an active
network formed by systems that have some intrinsic dynamics (e.g. neurons and
oscillators), how much information from such stimulus can be realized by
measuring the time evolution of one of the elements of the network ?
Determining how and how much information flows along anatomical brain paths is
an important requirement for the understanding of how animals perceive their
environment, learn and behave \cite{smith,eggermont,theunissen}.

The works of Refs. \cite{eggermont,theunissen,roland,smith,palus,dz} propose
ways to quantify how and how much information from a stimulus is transmitted
in neural networks. In particular, Ref. \cite{roland} demonstrated that 50$\%$
of the information about light displacements might be lost after being
processed by the H1 neuron, sensitive to image motion around a vertical axis,
a neuron localized in a small neural network of the Chrysomya magacephala fly,
the lobula plate. Does that mean that the H1 neuron has an information
capacity lower than the information contained in the light stimulus ?  Or does
that mean that information is lost due to the presence of internal noise ? 
These questions and others, which are still awaiting answers, concern the
rules under which information is coded and then transmitted by neurons and it
is a major topic of research in neuroscience referred to as the neural code
\cite{eggermont,theunissen}.

Even though the approaches of
Ref. \cite{eggermont,theunissen,roland,smith,palus,dz} have brought
considerable understanding on how and how much information from a stimulus is
transmitted in a neural network, the relation between network circuits
(topology) and information transmission in a neural as well as an active
network is still awaiting a more quantitative description
\cite{jirsa}. And that is the main thrust of the present manuscript, namely, to present
a quantitative way to relate network topology with information in active
networks. Since information might not always be easy to be measured or
quantified in experiments, we endevour to clarify the relation between information
and synchronization, a phenomenom which is often not only possible to observe
but also relatively easy to characterize.

We initially proceed along the same line as in Refs. \cite{schreiber,liang}, and
study the information transfer in autonomous systems. However, instead of
treating the information transfer between dynamical systems components, we
treat the transfer of information per unit time exchanged between two elements
in an autonomous chaotic active network. Thus, we neglect the complex relation
between external stimulus and the network and show how to calculate an upper
bound value for the mutual information rate (MIR) exchanged between two
elements (a communication channel) in an autonomous network. Ultimately, we
discuss how to extend this formula to non-chaotic networks suffering the
influence of a time-dependent stimulus.

Most of this work is directed to ensure the plausibility and validity of the
proposed formula for the upper bound of MIR (Sec. \ref{formula}) and also to
study its applications in order to clarify the relation among network
topology, information, and synchronization. We do not rely only on results
provided by this formula, but we also calculate the MIR by the methods in
Refs. \cite{baptista:2005,baptista:2007} and by symbolic encoding the
trajectory of the elements forming the network and then measuring the mutual
information provided by this discrete sequence of symbols (method described in
Sec. \ref{metodo1}).

To illustrate the power of the proposed formula, we applied it to study the
exchange of information in networks of coupled chaotic maps
(Sec. \ref{network_maps}) and in Hindmarsh-Rose neural networks
bidirectionally electrically coupled (Sec.\ref{network_HR}). The analyses are
carried out using quantities that we believe to be relevant to the treatment
of information transmission in active networks: a {\it communication channel},
the {\it channel capacity}, and the {\it network capacity} (see definitions in
Sec. \ref{metodo2}).

A communication channel represents a pathway through which information is
exchanged. In this work, a communication channel is considered to be formed by
a pair of elements. One element represents a transmiter and the other a
receiver, where the information about the transmiter can be measured.

The channel capacity is defined in terms of the proposed upper bound for the
MIR. It measures the local maximal rate of information that two elements in a
given network are able to exchange, a point-to-point measure of information
exchange. As we shall see, there are two network configurations for which
the value of the upper bound can be considered to be maximal with respect to
the coupling strength.

The network capacity is the maximum of the KS-entropy, for many possible
network configurations with a given number of elements. It gives the amount of
independent information that can be simultaneously transmitted within the
whole network, and naturally bounds the value of the MIR in the channels,
which concerns only the transmission of information between two elements.

While the channel capacity is bounded and does not depend on the number of
elements forming the network, the network capacity depends on the number of
elements forming the network.

As a direct application of the formula for the upper bound value of the MIR,
we show that an active network can operate with a large amount of MIR and
KS-entropy and at the same time it is robustly resistant to alterations in the
coupling strengths, if the eigenvalues of the Laplacian matrix satisfy some
specified conditions (Sec. \ref{eigenvalues}). The Laplacian matrix describes
the connections among the elements of the network.

The  conditions on the eigenvalues depend on whether the network is constructed in
order to possess communication channels that are either self-excitable or
non-self-excitable (see definition in Sec. \ref{metodo3}). Active networks
that possess non-self-excitable channels (formed by oscillators as the
R\"ossler, or the Chua's circuit) have channels that achieve their capacity
whenever their elements are in complete synchrony. Therefore, if a large
amount of information is desired to be transmitted point-to-point in a
non-self-excitable network, easily synchronizable networks are required.  On
the other hand, networks that possess self-excitable channels (as the ones
formed by neurons), achieve simultaneously its channel and network capacities
when there is at least one unstable mode of oscillation (time-scale) that is
out of synchrony (see Sec. \ref{metodo0}).

While non-self-excitable channels permit the exchanging of a moderate amount
of information in a reliable fashion, due to the low level of
desynchronization in the channel, self-excitable channels permit the exchange
of surprisingly large amounts of information, not necessary reliable, due to
the higher level of desynchronization in the channel.

In aiming at finding optimal network topologies, networks that can not only
transmit large amounts of information but are also robust under alterations
in the coupling strengths, we arrive at two relevant eigenvalues conditions
which provide networks that satisfy all the optimal requirements. Either the
network has elements that remain completely desynchronous for large variations
of the coupling strength, forming the self-excitable channels, or the network
has elements almost completely synchronous, forming the non-self-excitable
channels. In fact, the studied network, a network formed by electrically
connected Hindmarsh-Rose neurons, can have simultaneously self-excitable and
non-self-excitable channels.

Self-excitable networks, namely those that have a majority number of
self-excitable channels, have the topology of a perturbed star, i.e., they are
composed of a central neuron connected to most of the other outer neurons, and
some outer neurons sparsely connected among themselves. The networks that have
non-self-excitable channels have the topology of a perturbed fully 
connected network, i.e., a network whose elements are almost all-to-all
connected. The self-excitable network has thus a topology which can be
considered to be a model for mini-columnar structure of the mammalian
neocortex \cite{malsburg1}.

In order to construct optimal networks, we have used two approaches. Firstly,
(Sec. \ref{josue}), we use a Monte Carlo evolution technique
\cite{evorene} to find the topology of the network, assuming equal
bidirectional coupling strengths. This evolving technique simulates the
rewiring of a neuron network that maximizes or minimizes some cost function, in
this case a cost function which produces optimal networks to transmit
information. In the second approach (Sec. \ref{pecora}), we allow the elements
to be connected with different coupling strengths. We then use the Spectral
Theorem to calculate the coupling strengths of an all-to-all topology network.

Finally, we discuss how to extend these results to networks formed by elements
that are non-chaotic (Sec. \ref{NC}), and to non-autonomous networks, that are
being perturbed by some time-dependent stimuli (Secs. \ref{NC} and \ref{tds}).

\section{Results}

\section{Upper bound for the Mutual Information Rate (MIR) in an Active
Network}
\label{formula}

In a recent publication \cite{baptista:2005}, we have argued that the mutual
information rate (MIR) between two elements in an active chaotic network,
namely, the amount of information per unit time that can be realized in one
element, $k$, by measuring another element, $l$, regarded as $I_C$, is given
by the sum of the conditional Lyapunov exponents associated with the
synchronization manifold (regarded as $\lambda^{\parallel}$) minus the
positive conditional Lyapunov exponents associated with the transversal
manifold (regarded as $\lambda^{\perp}$). So, $I_C=\lambda^{\parallel} -
\lambda^{\perp}$. 

As shown in \cite{baptista:2007}, if one has N=2 coupled chaotic systems,
which produce at most two positive Lyapunov exponents $\lambda_1, \lambda_2$
with $\lambda_1>\lambda_2$, then $\lambda^{\parallel} = \lambda_1$ and
$\lambda^{\perp}=\lambda_2$. Denote the trajectory of the element $k$ in the
network by $\bf{x}_k$. For larger number of elements, $N$, the approaches
proposed in
\cite{baptista:2005} remain valid whenever the coordinate transformation
$\bf{X}_{kl \parallel}=\bf{x}_{k}+\bf{x}_{l}$ (which defines the
synchronization manifold) and $\bf{X}_{kl \perp}=\bf{x}_{k}-\bf{x}_{l}$ (which
defines the transversal manifold) successfully separates the two systems $k$
and $l$ from the whole network. Such a situation arises in networks of chaotic
maps of the interval connected by a diffusively (also known as electrically or
linear) all-to-all topology, where every element is connected to all the other
elements. These approaches were also shown to be approximately valid for
chaotic networks of oscillators connected by a diffusively all-to-all
topology. The purpose of the present work is to extend these approaches and
ideas to active networks with arbitrary topologies.

Consider an active network formed by $N$ equal elements, $\bf{x}_i$
($i=1,\ldots,N$), where every $D$-dimensional element has a different set of
initial conditions, i.e., ${\bf{x}}_1 \neq {\bf{x}}_2 \neq \ldots \neq
{\bf{x}}_N$. The network is described by
\begin{equation}
{\bf{\dot{x}}}_i={\bf F}({\bf{x}}_i)-\sigma \sum_j
{\bf{\mathcal{G}}}_{ij}{\bf{H}}({\bf{x}}_j),
\label{element_dynamics}
\end{equation}
where ${\bf{\mathcal{G}}}_{ij}$ is the $ij$ element of the coupling
matrix. Since we choose $\sum_j {\bf{\mathcal{G}}}_{ij}=0$ in order for a
synchronization manifold to exist by the subspace ${\bf \eta}={\bf x}_{1}={\bf
x}_2={\bf x}_3=\ldots={\bf x}_N$, we can call this matrix the Laplacian
matrix. The synchronous solution, ${\bf \eta}$, is described by
\begin{equation}
{\bf {\dot{\eta}}}=F({\bf{\eta}})
\label{synchronous}
\end{equation}

The way small perturbations propagate in the network
\cite{pecora} is described by the $i$ ($i=1,\ldots,N$) variational equations
of Eqs.  (\ref{element_dynamics}), namely writing ${\bf x_i}={\bf
\eta}+\delta{\bf x_i}$ and expanding Eq. (\ref{element_dynamics})
in $\delta {\bf x_i}$,  
\begin{equation}
\delta \dot{{\bf{x}}}_i=[\nabla {\bf F}({\bf{x}}_i)-
\sigma \sum_{j=1}^N {\bf{\mathcal{G}}}_{ij}D{\bf H}({\bf{x}}_i)]\delta {\bf{x}}_i
\label{variational}
\end{equation}
\noindent
obtained by linearly expanding Eq. (\ref{element_dynamics}). The spectra of Lyapunov
exponent is obtained from Eq. (\ref{variational}).
 
Making ${\bf{x}}_i=\xi$, which can be easily numerically done by setting the
elements with equal initial conditions and taking ${\bf H}({\bf x_i})={\bf
x_i}$, Eq. (\ref{variational}) can be made block diagonal resulting in
\begin{equation}
{\bf{\dot{\xi}}}_i=[\nabla {\bf F}({\bf{x}}_i) - \sigma \gamma_i ] {\bf \xi}_i.
\label{variational1}
\end{equation}
\noindent
where $\gamma_i$ are the eigenvalues (positive defined) of the Laplacian
matrix ordered such that $\gamma_{i+1} \ge \gamma_{i}$. Note that
$\gamma_1=0$.

Notice that the network dynamics is described by Eq. (\ref{element_dynamics}),
which assumes that every element has different initial conditions and
therefore different trajectories (except when the elements are completely
synchronized). On the other hand, Eq. (\ref{variational1}) that provides the
conditional exponents considers that all the initial conditions are equal.
The equations for ${\bf \xi}_1$ describe the propagation of perturbations on
the synchronization manifold ${\bf \xi}$, and the other equations describe
propagation of perturbations on the manifolds transversal to the
synchronization manifold. While Eq. (\ref{variational}) provides the set of
Lyapunov exponents of an attractor, Eq. (\ref{variational1}) provides the
Lyapunov exponents of the synchronization manifold and its transversal
directions.

Notice also that when dealing with linear dynamics, the Lyapunov exponents
[obtained from Eq. (\ref{variational})] are equal to the conditional exponents
[obtained from Eq. (\ref{variational1})] independently on the initial
conditions.

Then, the upper bound of the MIR that can be measured from an element
$\bf{x}_k$ by observing another element $\bf{x}_l$, i.e. the upper 
bound of the MIR in the communication channel $c^{i-1}$ is
\begin{equation}
I_P^{i-1} \leq |\lambda^1-\lambda^i|
\label{capacity}
\end{equation} 
\noindent
with $i \in (2,\ldots,N)$, and $\lambda^i$ representing the sum of all the
{\bf positive} Lyapunov exponents of the equation for the mode $\xi_i$, in
Eq. (\ref{variational1}). So, $\lambda^1$ is the sum of the positive
conditional exponents obtained from the separated variational equations, using
the smallest eigenvalue associated with the exponential divergence between
nearby trajectories around $\xi$, the synchronous state, and $\lambda^i$
($i>1$) are the sum of the positive conditional exponents of one of the
possible desynchronous oscillation modes. Each eigenvalue $\gamma_i$ produces
a set of conditional exponents $\lambda^i_m$, with $m=1,\ldots,D$.

Each oscillatory mode $\xi_i$ represents a subnetwork within the whole network
which possesses some oscillatory pattern. This oscillatory subnetwork can be
used for communications purposes. Each mode represents a path along which
information can be transmitted through the network. The oscillation mode
associated with the synchronization manifold ($\xi_1$) propagates some
information signal everywhere within the network. The desynchronous modes
limits the amount of information that one can measure from the signal
propagated by the synchronous mode. Although Eq. (\ref{capacity}) gives the
upper bound for the amount of information between modes of oscillation, for
some simple network geometries, as the ones studied here, we can relate the
amount of information exchanged between two vibrational modes to the amount of
information between two elements of the network, and therefore,
Eq. (\ref{capacity}) can be used to calculate an upper bound for the MIR
exchanged between pairs of elements in the network. For larger and complex
networks, this association is non-trivial, and we rely on the reasonable
argument that a pair of elements in an active network cannot transmit more
information than some of the $i-1$ values of $I_P^{i-1}$.

The inequality in Eq. (\ref{capacity}) can be interpreted in the
following way. The right hand side of Eq.  (\ref{capacity}) calculates
the amount of information that one could transmit if the whole network
were completely synchronous with the state $\xi$, which is only true
when complete synchronization takes place. Typically, we expect that
the elements of the network will not be completely synchronous to
$\xi$.  While the positive conditional exponents associated with the
synchronization manifold provide an upper bound for the rate of
information to be transmitted, the transversal conditional exponents
provide a lower bound for the rate of erroneously information
exchanged between nodes of the network. Thus, the amount of
information provided by the right part of Eq.  (\ref{capacity})
overestimates the exact MIR which, due to desynchronization in the
network, should be smaller than the calculated one. For more details
on the derivations of Eq. (\ref{capacity}), see
Sec. \ref{network_maps}.

Equation (\ref{rescale}) allows one to calculate the MIR between oscillation
modes of larger networks with arbitrary topology rescaling the MIR curve
($I_P^1$ vs. $\sigma$) obtained from two coupled elements. Denoting
$\sigma^{*}(N=2)$ as the strength value for which the curve for $\lambda^2$
reaches a relevant value, say, its maximum value, then the coupling strength
for which this same maximum is reached for $\lambda^i$ in a network composed
by $N$ elements is given by
\begin{equation}
\sigma^{i*}(N)=\frac{2\sigma^{*}(N=2)}{\gamma_i(N)}
\label{rescale}
\end{equation}
\noindent
where $\gamma_i(N)$ represents the $i$th largest eigenvalue of the
$N$-elements network. If the network has an all-to-all topology, thus,
$\sigma^{*}(N=2)$ represents the strength value for which the curve of $I_P^1$
reaches a relevant value, and $\sigma^{*}(N)$ the strength value that this
same value for $I_P^i$ is reached.

Notice that symmetries in the connecting network topology leads to the presence
of degenerate eigenvalues (=equal eigenvalues) in the Laplacian matrix, which
means that there are less independent channels of communication along which
information flows.  Calling $Q$ the number of degenerate eigenvalues of the
Laplacian matrix, Eq. (\ref{capacity}) will provide $N-Q$ different values.

As the coupling strength $\sigma$ is varied, the quantities that measure
information change correspondingly. For practical reasons, it is important
that we can link the way these quantities (see Sec. \ref{metodo2}) change with
the way the different types of synchronization show up in the network (see
Sec. \ref{metodo0}). In short, there are three main types of synchronization
observed in our examples: burst phase synchronization (BPS), when at least one
pair of neurons are synchronous in the slow time-scale but desynchronous in
the fast time-scale, phase synchronization (PS), when all pairs of neurons are
phase synchronous, and complete synchronization (CS), when all pairs of
neurons are completely synchronous. The coupling strength for which these
synchronous phenomena appear are denoted by $\sigma_{BPS}$, $\sigma_{PS}$, and
$\sigma_{CS}$ (with no superscript index).

Finally, there are a few more relevant coupling strengths, which characterize
each communication channel. First, $\sigma_{min}^{i}$, for which the sum of
the $i$th conditional exponents $\lambda^i$ equals the value of $\lambda^1$. For
$\sigma<\sigma_{min}^{i}$, the communication channel $i$ (whose upper rate of
information transmission depends on the two oscillation modes $\xi_1$ and
$\xi_i$) behaves in a self-excitable way, i.e., $\lambda^1 <
\lambda^i$. For $\sigma \geq \sigma_{min}^{i}$, 
$\lambda^1 \geq \lambda^i$. Secondly, $\sigma^{i*}$ indicates the coupling
strength at which $I_P^{i-1}$ is maximal. Thirdly, $\sigma^{i}_{CS}$ indicates
the coupling strength for which the communication channel $c^{i-1}$ becomes
"stable", i.e., $\lambda^{i} < 0$. At $\sigma=\sigma^{i*}$ the self-excitable
channel capacity of the channel $c^{i-1}$ is reached and at
$\sigma=\sigma^{i}_{CS}$, the non-self-excitable channel capacity is
reached. Finally, $\sigma_{C}$ is the coupling for which the network capacity
is reached, and then, when the KS-entropy of the network is maximal. For other
quantities, see Sec. \ref{metodo2}.

\section{The MIR in networks of coupled Hindmarsh-Rose neurons}\label{network_HR}

We investigate how information is transmitted in self-excitable
networks composed of $N$ bidirectionally coupled Hindmarsh-Rose neurons
\cite{hindmarsh}:
\begin{eqnarray}
\dot{x}_i &=& y_i + 3x_i^2 - x_i^3-z_i + I_i +
\sigma \sum_j {\mathcal{G}}_{ij}(x_j) \nonumber \\
\dot{y}_i &=& 1-5x_i^2-y_i \label{HR} \\
\dot{z}_i  &=& -rz_i + 4r(x_i+1.6) \nonumber
\end{eqnarray}
\noindent
The parameter $r$ modulates the slow dynamics and is set equal to 0.005, such
that each neuron is chaotic.  The index $i \neq j$ assumes values within the
set $[1,\ldots,N]$.  $S_k$ represents the subsystem formed by the variables
$(x_k,y_k,z_k)$ and $S_l$ represents the subsystem formed by the variables
$(x_l,y_l,z_l)$, where $k$=$[1,\ldots,N-1]$ and $l$=$[k+1,\ldots,N]$.  The
Laplacian matrix is symmetric, so ${\bf \mathcal{G}}_{ji}={\bf
\mathcal{G}}_{ij}$, and $\sigma {\bf \mathcal{G}}_{ji}$ is the strength of the
electrical coupling between the neurons, and we take for $I_i$ the value 
$I_i=3.25$.

In order to simulate the neuron network and to calculate the Lyapunov
exponents through Eq. (\ref{variational}), we use the initial conditions
$x$=-1.3078+$\eta$, $y$=-7.3218+$\eta$, and $z$=3.3530+$\eta$, where $\eta$ is
an uniform random number within [0,0.02]. To calculate the conditional
Lyapunov exponents, we use the equal initial conditions, $x$=-1.3078,
$y$=-7.3218, and $z$=3.3530.

\begin{figure}[!htb]
\centerline{\hbox{\psfig{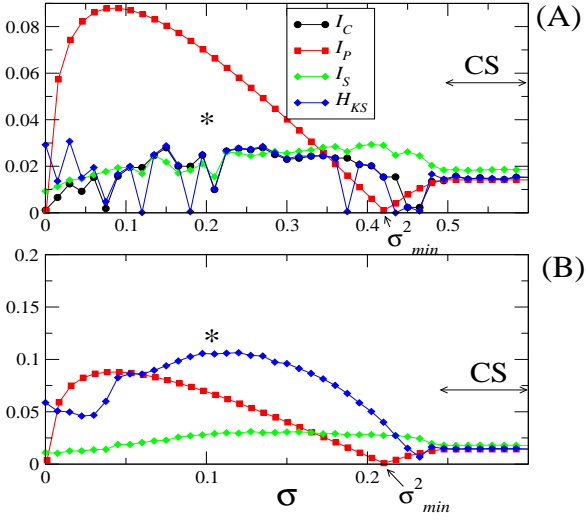}}}
\caption{The quantities $I_C$ (black circles), $I_P$ (red squares), 
$I_S$ (green diamonds), and $H_{KS}$ (blue diamonds), for two (A) and four (B)
coupled neurons, in an all-to-all topology. Notice that since there are only
two different eigenvalues, there is only one channel of communication whose
upper bound for the MIR is given by $I_P=|\lambda^1-\lambda^2|$. Also, $I_S$
and $I_C$ represent the mutual information exchanged between any two pairs of
elements in the system. In (A), $\sigma^{2*}$=0.092, $\sigma_{BPS} \cong 0.2$,
$\sigma^{2}_{min}=$0.42, $\sigma_{PS}=0.47$, and $\sigma_{CS}$=0.5. In (B),
$\sigma^{2*}$=0.046, $\sigma_{BPS} \cong 0.1$, $\sigma^2_{min}=$0.21,
$\sigma_{PS}=0.24$, and $\sigma_{CS}$=0.25. CS indicates the coupling interval
$\sigma \geq \sigma_{CS}$ for which there exists complete synchronization.}
\label{DD_fig1}
\end{figure}

{\bf All-to-all coupling:} Here, we analyze the case where $N$ neurons are
fully connected to every other neuron. The Laplacian matrix has $N$
eigenvalues, $\gamma_1$=0, and $N-1$ degenerate ones $\gamma_i$=$N$,
$i=2,\ldots,N$.  Every pair of neurons exchange an equal amount of
MIR. Although, there are $N\times(N-1)/2$ pairs of neurons, there is actually
only one independent channel of communication, i.e., a perturbation applied at
some point of the network should be equally propagated to all other points in
the network. In Fig. \ref{DD_fig1}(A), we show the MIR, $I_C$, calculated
using the approaches in Refs.
\cite{baptista:2005,baptista:2007}, $I_P$, calculated using the right hand-side
of Eq. (\ref{capacity}), and $I_S$, calculated encoding the trajectory between
pair of neurons (Sec. \ref{metodo1}), and the Kolmogorov-Sinai entropy,
$H_{KS}$, for a network composed by $N$=2 neurons. In (B), we show these same
quantities for a network formed by $N$=4 neurons.

While for $\sigma\cong 0$ and $\sigma \geq \sigma_{CS}$, we have that $I_C
\cong I_P \cong I_S$, for $\sigma \cong \sigma^{2*}$ (when the self-excitable
channel capacity is reached) it is clear that $I_P$ should be an upper bound
for the MIR, since not only $I_P>I_C$ but also $I_P>I_S$. Notice the good
agreement between $I_C$ and $I_S$, except for $\sigma\cong \sigma^2_{min}$,
when $I_S>H_{KS}$, which violates Eq. (\ref{condicaoI_H_KS}).

The star symbol indicates the value of the coupling, $\sigma_{BPS}$
(Sec. \ref{metodo0}), for which burst phase synchronization (BPS) appears
while the spikes are highly desynchronous. The appearance of BPS coincides
with the moment where all the quantifiers for the MIR are large, and close to
a coupling strength, $\sigma_C$, for which the network capacity is reached
(when $H_{KS}$ is maximal).

At this point, the network is sufficiently desynchronous to generate a large
amount of entropy, which implies a large $\lambda^i$, for $i \geq 2$.  This
is an optimal configuration for the maximization of the MIR.  There exists
phase synchrony in the subspace of the slow time-scale $z$ variables (which is
responsible for the bursting-spiking behavior), but there is no synchrony in
the $(x,y)$ subspace. This supports the binding hypothesis, a fundamental
concept of neurobiology
\cite{malsburg1} which sustains that 
 neural networks coding the same feature or object are functionally bounded.
 It also simultaneously supports the works of
\cite{pareti}, which show that desynchronization seems to play an
important role in the perception of objects as well. Whenever $\lambda^2$
approaches zero, at $\sigma=\sigma_{CS}$, there is a drastic reduction in
the value of $H_{KS}$ as well as $I_P$, since the network is in complete
synchronization (CS), when all the variables of one neuron equals the
variables of the other neurons. 

Therefore, for coupling strengths larger than the one indicated by the star
symbol, and smaller than the one where CS takes place, there is still one
time-scale, the fast time-scale, which is out of synchrony.

For $\sigma > \sigma^2_{min}$, the only independent communication channel is of
the non-self-excitable type. That means $\lambda^i \leq \lambda^1$ ($i \geq
2$), and as the coupling strength increases, $H_{KS}$ decreases and $I_P$
increases.

Note that the curve for $I_P$ shown in Fig. \ref{DD_fig1}(B) can be obtained
by rescaling the curve shown in Fig. \ref{DD_fig1}(A), applying
Eq. (\ref{rescale}).

\begin{figure}[!htb]
\centerline{\hbox{\psfig{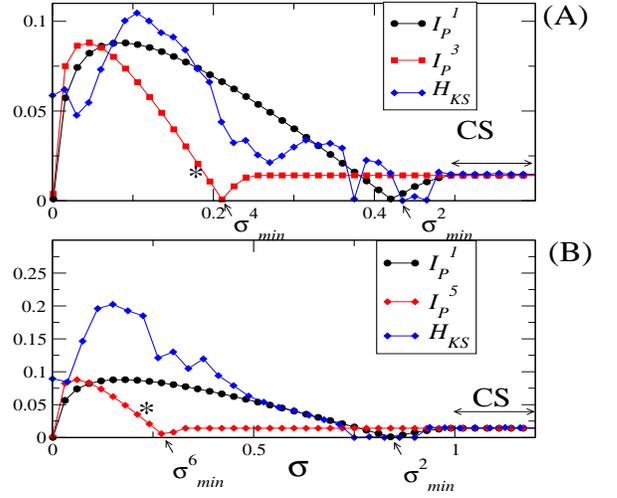}}}
\caption{The quantities $I_P$ and $H_{KS}$ for 
nearest-neighbor networks with $N$=4 (A) and $N$=6 (B). In (A),
$\sigma^{2*}$=0.09, $\sigma^{4*}$=0.046, $\sigma_{min}^2$=0.42,
$\sigma_{min}^4$=0.21, $\sigma^4_{CS}$=0.25,
$\sigma_{BPS} \cong 0.18$, $\sigma_{PS}$=0.462, and $\sigma_{CS}$=0.5. In (B),
$\sigma^{2*}$=0.18, $\sigma^{6*}$=0.061, $\sigma_{min}^2$=0.84,
$\sigma_{min}^6$=0.27, $\sigma^6_{CS}$=0.33, $\sigma_{BPS}
\cong 0.23$, $\sigma_{PS}$=0.78, and $\sigma_{CS}$=1.0. The stars point to
where BPS first appears. CS indicates the coupling interval $\sigma \geq
\sigma_{CS}$ for which there exists complete synchronization.}
\label{DD_fig2}
\end{figure}

{\bf Nearest-neighbor coupling:} Here, every neuron is connected to its
nearest neighbors, with periodic boundary conditions, forming a closed
ring. The eigenvalues of the Laplacian matrix can be calculated from
$\gamma_{k}$=$4\sin{(\frac{\pi (k-1)}{N})}^2$, $k
\in [1,\ldots,N]$. Notice that in this example, $\gamma_{k+1}$ might be smaller
than $\gamma_{k}$ due to the degeneracies. We organize the eigenvalues
in a crescent order. For our further examples, we consider $N$=4 [in Fig. 
\ref{DD_fig2}(A)] and $N$=6 [in Fig. \ref{DD_fig2}(B)]. For $N$=4,
$\gamma_1$=0, $\gamma_{2,3}$=2, $\gamma_{4}$=4, and for $N$=6, $\gamma_1$=0,
$\gamma_{2,3}$=1, $\gamma_{4,5}$=3, $\gamma_6$=4.

Networks with a nearest-neighbor coupling topology and an even number of
elements possess a connecting matrix ${\bf \mathcal{G}}$ with $N/2-1$
degenerate eigenvalues, and therefore, $N-N/2+1$ distinct eigenvalues. There
are only $N-N/2$ different minimal path lengths connecting the elements of the
network. The minimal path length quantifies the minimal distance between an
element and another in the network by following a path formed by connected
elements. Note that $I_P$ assumes only $N-N/2$ different values. It is
reasonable to state that each different value corresponds to the exchange of
information between elements that have the same minimal path length.

For a network with $N$=4 [Fig. \ref{DD_fig2}(A)], there are two possible
minimal path lengths, 1 and 2. Either the elements are 1 connection apart, or
2 connections apart. For such a network, it is reasonable to associate $I_P^1
=
\lambda^1-\lambda^2$ with the MIR between two elements, $S_k$ and $S_{k+2}$, that are 
2 connections apart, and $I_P^3 = |\lambda^1-\lambda^4|$ to the MIR between
two elements, $S_k$ and $S_{k+1}$, that are 1 connection apart.  The more
distant (closer) an element is from any other, the larger (smaller) the
coupling strength for them to synchronize. In addition,
$\sigma^{2*}>\sigma^{4*}$ and $\sigma^2_{min}>\sigma^4_{min}$. That means that
the more distant elements are from each other the larger the coupling strength
is, in order for these two elements to exchange a large rate of information, since
$\sigma^{2*}>\sigma^{4*}$. In addition, since $\sigma^2_{min}>\sigma^4_{min}$,
the communication channel responsible for the exchange of information between
closer elements (the channel $c^3$) becomes non-self-excitable for a smaller
value of the coupling strength than the strength necessary to turn the
communication channel responsible for the exchange of information between
distant elements (the channel $c^1$) into a non-self-excitable channel. Since
the level of desynchronization in a non-self-excitable channel is low, then,
closer elements can exchange reliable information for smaller coupling
strengths than the strength necessary for distant elements to exchange
reliable information. Note that due to the 1 degenerated eigenvalue,
$I_P^1$=$I_P^2$, $\sigma^{2*}=\sigma^{3*}$, and
$\sigma^2_{min}=\sigma^3_{min}$. A similar analysis can be done for the
network $N$=6, whose results are shown in Fig. \ref{DD_fig2}(B).
 
The KS entropy of the network, $H_{KS}$, is also shown in this figure.  In
(A), $\sigma_{min}^2$=0.42 and $\sigma_{min}^4$=0.21, and in (B),
$\sigma_{min}^2$=0.84 and $\sigma_{min}^6$=0.275, values that can be easily
derived from Eq. (\ref{rescale}). Note that the values of
$\sigma=\sigma_{min}^4$ in (A) [and $\sigma=\sigma_{min}^6$, in (B)] are close
to the parameter for which BPS in the slow time-scale is first observed in
these networks (indicated by the star symbol in Fig. \ref{DD_fig2}),
$\sigma_{BPS}
\cong 0.18$ [in (A)] and $\sigma_{BPS} \cong 0.23$ [in (B)]. 
At $\sigma \cong \sigma_{min}^4$ [in (A)] and $\sigma \cong \sigma_{min}^6$
[in (B)], also the quantities $I_P^1$ and $H_{KS}$ are large.

Another important point to be emphasized in these networks is that $\Delta
\sigma^i_{NSE}$ = $\sigma_{CS}-\sigma_{min}^i$, regarded as the 
non-self-excitable robustness parameter for the communication channels $c^i$,
with $i$=3 for the network with $N$=4 [in (A)] and $i=5$ for the network with
$N$=6 [in (B)] is large.  This is a consequence of the fact that the
normalized spectral distance (NED), $(\gamma_i-\gamma_2)/N$ is also large, for
either $i=4$ [in (A)] or $i=6$ [in (B)]. Having a large NED between the $i$th
largest and the first largest eigenvalues results in a non-self-excitable
channel, $c^{i-1}$, robust under large alterations of the coupling
strength. On the other hand, $\Delta
\sigma^i_{SE}$ = $\sigma^i_{min}$, regarded as the 
self-excitable robustness parameter for the communication channel $c^{i-1}$,
is large, for $i=2,3$. This is a consequence of the fact that the normalized
spectral distance (NED), $(\gamma_N-\gamma_{i})/N$ is large.  Having a large
NED between the largest and the $i$th largest eigenvalues results in a
self-excitable channel, $c^{i-1}$, robust under large alterations of the
coupling strength.

Notice also that the maximal values of $I_P$ for the all-to-all and
nearest-neighbor networks topologies is the same (see Figs. \ref{DD_fig1} and
\ref{DD_fig2}). This shows that the maximum of $I_P$ does not depend on the
number, 
$N$, of elements in the network. Not so in the case of the network capacity
${\mathcal{C}}_C$, which increases with $N$. Thus, pairs of elements
can transmit information in a rather limited rate, but depending on the number
of elements forming the network, a large number of channels can simultaneously
transmit information.

{\bf Star coupling: } We consider $N$=4. There is a central neuron, denoted by $S_1$,
bidirectionally connected to the other three ($S_k, k=2,3,4$), but
none of the others are connected among themselves.  The eigenvalues of the
Laplacian matrix are $\gamma_1$=0,$\gamma_{2,3}$=1,$\gamma_4=N$. So, not only
the NED between $\gamma_N$ and $\gamma_{N-1}$ is large but also between
$\gamma_N$ and $\gamma_{N-2}$, and therefore, $\Delta
\sigma^{N-1}_{SE}$ and $\Delta
\sigma^{N-2}_{SE}$ are large.   
This provides a network whose channels $c^1$ and $c^2$ have a large MIR for a
large coupling strength alteration. Note that if $\gamma_{N-1}$ is far away
from $\gamma_N$ that implies that $\gamma_{N-2}$ is also far away from
$\gamma_{N}$. Thus, a reasonable spectral distance between $\gamma_{N-1}$ and
$\gamma_N$ is a ``biological requirement'' for the proper function of the
network, since even for larger coupling strengths there will be at least one
oscillation mode which is desynchronous, a configuration that enables
perturbation (meaning external stimuli) to be propagated within the network
\cite{diseases}.

\begin{figure}[!htb]
  \centerline{\hbox{\psfig{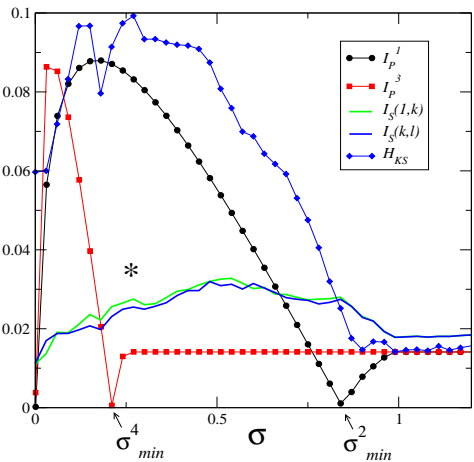}}}
\caption{MIR between the central neuron and an outer one (black circles),
$I_P^1$, (resp. $I_S(1,k)$, in green line), and between two outer ones (red
squares), $I_P^3$, (resp. $I_S(k,l)$, in blue line). Blue diamonds represents
the KS-entropy. Other quantities are $\sigma^{4*}=0.181$, $\sigma^{2*}=0.044$,
$\sigma^4_{min}=0.84$, $\sigma^2_{min}=0.22$,
$\sigma^4_{CS}$=0.27, $\sigma_{BPS}$=0.265,
$\sigma_{PS}$=0.92, and $\sigma_{CS}$=1.0. The star indicates the parameter for
which BPS first appears.}
\label{DD_fig3}
\end{figure}

The largest eigenvalue is related to an oscillation mode where all the outer
neurons are in synchrony with each other but desynchronous with the central
neuron. So, here it is clear the association between $|\lambda^1-\lambda^4|$ and
the MIR between the central neuron with an outer neuron, since
$\lambda^1$ represents the amount of information of the synchronous
trajectories among all the neurons, while $\lambda^4$ is the amount of
information of the desynchronous trajectories between the
central neuron and any outer neuron. The other
eigenvalues ($\gamma_2$,$\gamma_3)$ represent directions transverse to the
synchronization manifold in which the outer neurons become desynchronous with
the central neuron in waves wrapping commensurately around the central neuron
\cite{pecora}. Thus, $\lambda^2$ and $\lambda^3$ are related to the
error in the transmission between two outer neurons, $k$ and $l$, with $k,l
\neq 1$. 

Note that the MIR between $S_1$ and an outer neuron (upper bound represented
by $I_P^3=|\lambda^1-\lambda^4|$ and $I_S$ represented by $I_S(1,k)$, in
Fig. \ref{DD_fig3}) is larger (smaller) than the MIR between two outer neurons (upper
bound represented by $I_P^1=|\lambda^1-\lambda^2|$ and $I_S$ is represented by
$I_S(k,l)$, in Fig. \ref{DD_fig3}), for small coupling (for when the
channel $c^3$ is self-excitable, and $\sigma \geq \sigma^4_{min}$). Similar to the
nearest-neighbor networks, the self-excitable and the non-self-excitable
channel capacities of the channel associated with the transmission of
information between closer elements (the channel $c^3$) are achieved for a
smaller value of the coupling strength than the one necessary to make the
channels associated with the transmission of information between more distant
elements (the channel $c^1$) to achieve its two channel capacities.  That
property permits this network, for $\sigma \cong \sigma^4_{min}$, to transmits
simultaneously reliable information using the channel $c^3$ and with a higher
rate using the channel $c^1$.

Notice, in Fig.
\ref{DD_fig3}, that $\sigma^{2*} \cong \sigma^4_{min} \cong \sigma_{BPS} \cong
\sigma_C$. So, 
when the channel capacity of the channel $c^1$ is reached, also $H_{KS}$ of
the network is maximal, and the network operates with its capacity.

Another point that we want to emphasize in this network is that while a large
NED between $\gamma_N$ and $\gamma_{N-1}$ provides a network whose channel
$c^1$ is self-excitable and can transmit information at a large rate for a
large coupling strength interval, a large NED between $\gamma_3$ and
$\gamma_2$ leads to a non-self-excitable channel $c^3$ even for small values
of the coupling amplitudes, and it remains non-self-excitable for a large
variation of the coupling strength.  Thus, while a large NED between the
second and the first largest eigenvalues leads to a network whose channels are
predominantly of the self-excitable types, a large NED between the second
largest and the third largest eigenvalues provide a network whose
communication channels are predominantly of the non-self-excitable types.

\section{Eigenvalues conditions for optimal network topologies}\label{eigenvalues}

Finding network topologies and coupling strengths in order to have a network
that operates in a desired fashion is not a trivial task (see
Sec. \ref{metodo4} and \ref{metodo5}). An ideal way to proceed would be to
evolve the network topology in order to achieve some desired behavior. In this
paper, we are interested in maximizing simultaneously $I_P$, the KS-entropy,
and the average $\langle I_P \rangle$, for a large range of the coupling
strength, characteristics of an {\it optimal network}. However, evolving a
network in order to find an optimal one would require the calculation of the
MIR in every communication channel and $H_{KS}$ for every evolution step.
For a typical evolution, which requires 10$^6$ evolution steps, such an
approach is impractical.

Based on our previous discussions, however, an optimal network topology can be
realized by only selecting an appropriate set of eigenvalues which have some
specific NED. Evolving a network by the methods of Secs.
\ref{metodo4} and \ref{metodo5} using a cost function which is a function of 
only the eigenvalues of the Laplacian matrix is a practical and physible task.

The present section is dedicated to describe the derivation of this cost
function. 

\begin{figure}[!htb]
  \centerline{\hbox{\psfig{file=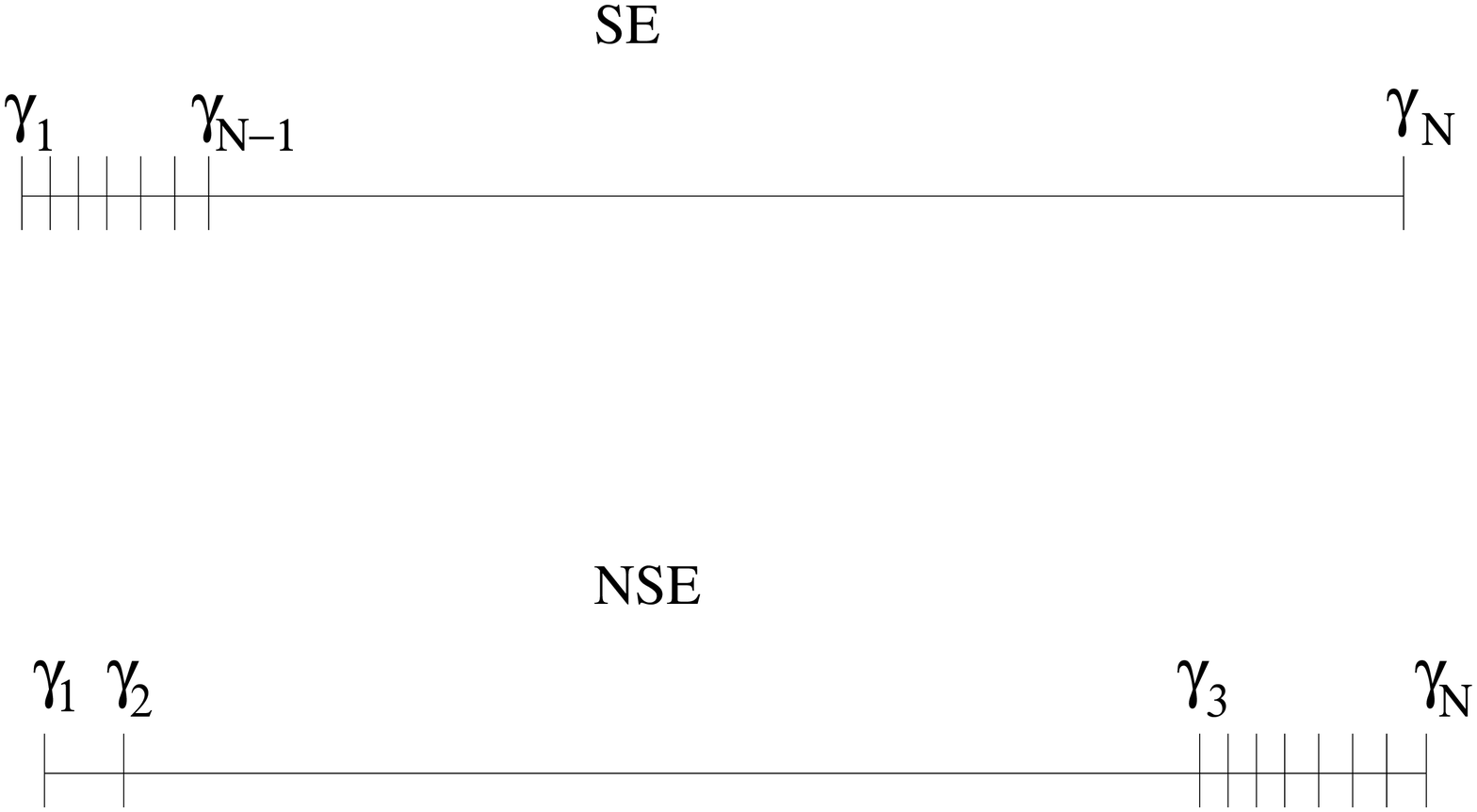,height=8cm,width=8cm,viewport=0
  -15 650 400}}}
\caption{Representation of the eigenvalues sets that produce optimal self-excitable (SE) and non-self-excitable active networks (NSE).}
\label{eigenvalues_set}
\end{figure}

We can think of two most relevant sets of eigenvalues which create optimal
networks, and they are represented in Fig. \ref{eigenvalues_set}. Either it is
desired eigenvalues that produce a network predominantly self-excitable
[SE, in Fig. \ref{eigenvalues_set}] or predominantly non-self-excitable [NSE,
in Fig. \ref{eigenvalues_set}].

In a network whose communication channels are predominantly self-excitable, it
is required that the NED $(\gamma_{N}-\gamma_{N-1})/N$ is maximal and
$(\gamma_{N-1})/N$ minimal. Therefore, we want a network for which the cost
function
\begin{equation}
{\mathcal{B}}_1 \equiv \frac{\gamma_N-\gamma_{N-1}}{\gamma_{N-1}}
\label{cond1}
\end{equation}
\noindent
is maximal. 

A network whose eigenvalues maximize ${\mathcal{B}}_1$  has
self-excitable channels for a large variation of the coupling strength.  As a
consequence, $\langle I_P \rangle$ as well as $H_{KS}$ is large for $\sigma
\in [\sigma^N_{min}, \sigma^2_{min}]$. 

In a network whose communication channels are predominantly
non-self-excitable, it is required that the NED $(\gamma_{3}-\gamma_2)/N$ is
maximal and $(\gamma_2)/N$ minimal. Therefore, we want a network for which the
cost function
\begin{equation}
{\mathcal{B}}_2 \equiv \frac{\gamma_3-\gamma_{2}}{\gamma_{2}}
\label{cond2}
\end{equation}
\noindent
is maximal.

A network whose eigenvalues maximize the condition in Eq. (\ref{cond2}) have
non-self-excitable channels for a large variation of the coupling strength.
As a consequence, $\langle I_P \rangle$ is large for $\sigma \in
[\sigma^N_{min}, \sigma^3_{min}]$, which is a small coupling range, but since
there is still one oscillation mode that is unstable (the mode $\xi_2$),
$H_{KS}$ is still large for a large range of the coupling strength ($\sigma <
\sigma^2_{min}$). Most of the channels will transmit information in a reliable
way, since the error in the transmission, provided by $\lambda^i$ ($i \geq
2$), of most of the channels will be zero, once $\lambda^i<0$.

Since degenerate eigenvalues produce networks with less vibrational modes, we
assume in the following the absence of such degenerate eigenvalues. In
addition, we assume that there is a finite distance between eigenvalues so
that the network becomes robust under rewiring, and therefore, perturbing
${\mathcal{G}}_{ij}$ will not easily create degenerate eigenvalues.

A network that is completely synchronous and has no unstable modes does not
provide an appropriate environment for the transmission of information about
an external stimulus, because they prevent the propagation of perturbations.
Networks that can be easily completely synchronized (for small coupling
strengths) requires the minimization of $\gamma_N-\gamma_2$, or in terms of
the eigenratio, the minimization of $\gamma_N/\gamma_2$.  We are not
interested in such a case. To construct network topologies that
are good for complete synchronization, see Refs.
\cite{pecora,stefano,jurgen}.

\section{Optimal topologies for information transmission}

Before explaining how we obtain optimal network topologies for information
transmission, it is important to discuss the type of topology expected
to be found by maximizing either ${\mathcal{B}}_1$, in Eq. (\ref{cond1}) or
${\mathcal{B}}_2$, in Eq. (\ref{cond2}).  Notice that Laplacians
whose eigenvalues maximize ${\mathcal{B}}_1$ are a perturbed version of the
star topology, and the ones that maximize ${\mathcal{B}}_2$ are a perturbed
version of the all-to-all topology.  In addition, in order to have a network
that presents many independent modes of oscillations it is required that the
Laplacian matrix presents as much as possible, a large number of
non-degenerate eigenvalues.  That can be arranged by rewiring (perturbing)
networks possessing either the star or the nearest-neighbor topology, breaking
the symmetry.

In order to calculate an optimal Laplacian, we propose two approaches.

One approach, described in Sec. \ref{metodo4}, is based on the reconstruction
of the network by evolving techniques, simulating the process responsible for
the growing or rewiring of real biological networks, a process which tries to
maximize or minimize some cost function. The results are discussed in Sec.
\ref{josue}.

A second approach, described in Sec. \ref{metodo5}, is based on the Spectral
Theorem, and produces a network in order for its Laplacian to have a
previously chosen set of eigenvalues. These eigenvalues can be chosen in order
to maximize the cost function. The results are discussed in Sec.
\ref{pecora}.

\subsection{Evolving networks}\label{josue}

\begin{figure}[!htb]
  \centerline{\hbox{\psfig{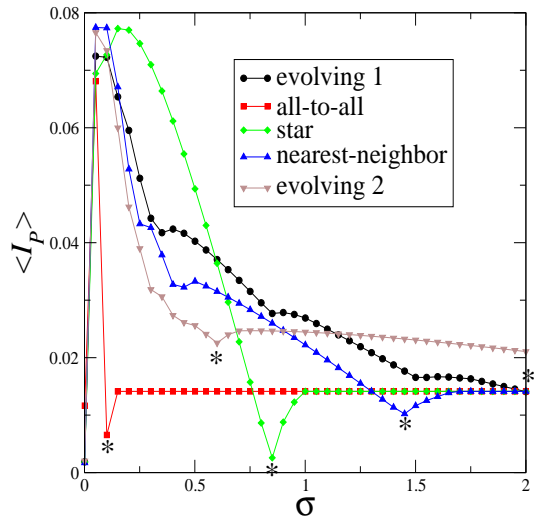}}}
\caption{The average value of the upper bound MIR, $\langle I_P \rangle$ [as defined in
Eq. (\ref{averagedIP})] for active networks composed of 8 elements using one
of the many topologies obtained by evolving the network maximizing
${\mathcal{B}}_1$ (circles), all-to-all topology (squares), star topology
(diamonds), nearest-neighbor (upper triangle), and maximizing
${\mathcal{B}}_2$ (down triangle). The values of $\sigma^2_{min}$ indicated by
the starts are $\sigma^2_{min}$=0.169 (evolving 1), $\sigma^2_{min}$=0.05
(all-to-all), $\sigma^2_{min}$=0.037 (star), $\sigma^2_{min}$=0.037
(nearest-neighbor), and $\sigma^2_{min}$=0.6 (evolving 2).
The evolving 1 network has a Laplacian with relevant eigenvalues
$\gamma_{7}$=3.0000, $\gamma_{8}$=6.1004, which produces
a cost function equal to ${\mathcal{B}}_1$=1.033. The evolving 2 network has a
Laplacian with relevant eigenvalues $\gamma_2$=0.2243 and $\gamma_{3}$=1.4107,
which produces a cost function equal to ${\mathcal{B}}_2$=5.2893.}
\label{arbitrary_fig02}
\end{figure}

In order to better understand how a network evolves (grows) in accordance with
the maximization of the cost functions in Eqs. (\ref{cond1}) and
(\ref{cond2}), we first find the network configurations with a small
number of elements. To be specific, we choose $N$=8 elements. To show that
indeed the calculated network topologies produce active networks that operate
as desired, we calculate the average upper bound value of the MIR
[Eq. (\ref{averagedIP})] for neural networks described by Eqs. (\ref{HR}) with
the topology obtained by the evolution technique, and compare with other
network topologies. Figure
\ref{arbitrary_fig02} shows $\langle I_P \rangle$, the average channel
capacity, calculated for networks composed of 8 elements, using one of the
many topologies obtained by evolving the network maximizing ${\mathcal{B}}_1$
(circles, denoted in Fig. by "evolving 1"), all-to-all topology (squares),
star topology (diamonds), nearest-neighbor (upper triangle), and maximizing
${\mathcal{B}}_2$ (down triangle, denoted in Fig. by "evolving 2").  The star
points to the value of $\sigma^2_{min}$, when $c^1$, the most unstable
communication channel (a self-excitable channel), becomes non-self-excitable.

As desired the evolving network 1 has a large upper bound for the MIR (as
measured by $\langle I_P \rangle$) for a large range of the coupling strength,
since the network has predominantly self-excitable channels. The channel $c^1$
has a large robustness parameter $\Delta\sigma^2_{SE}$, i.e., it is a
self-excitable channel for $\sigma<\sigma^{2}_{min}$, where
$\sigma^{2}_{min}$=2.0.  In contrast to the other topologies, in the star,
nearest-neighbor, and all-to-all topologies, $\Delta \sigma_{SE}^2$ is smaller
and $\Delta \sigma_{NSE}^2$ is larger.  Even though most of the channels in
the evolving 2 topology are of the non-self-excitable type, $\langle I_P
\rangle$ remains large even for higher values of the coupling 
strength. That is due to the channel $c^1$ which turns into a self-excitable
channel only for $\sigma>2$.

\begin{figure}[!htb]
  \centerline{\hbox{\psfig{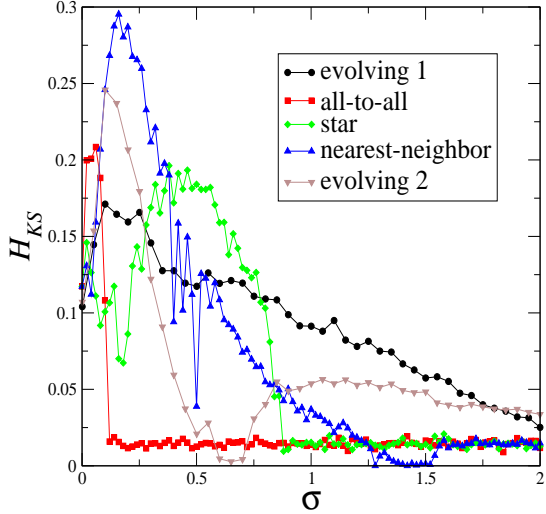}}}
\caption{KS-entropy for the same active networks of Fig. \ref{arbitrary_fig02}
 composed of 8 elements.}
\label{arbitrary_fig03}
\end{figure}

The KS-entropies of the 5 active networks whose $\langle I_P \rangle $ are
shown in Fig. \ref{arbitrary_fig02} are shown in
Fig. \ref{arbitrary_fig03}. Typically, the network capacities are reached for
roughly the same coupling strength for which the maximum of $\langle I_P
\rangle$, is reached.  
In between the coupling strength for which the network capacities and the
maximal of $\langle I_P \rangle$ are reached, $\lambda^3$ becomes negative. At
this point, also BPS appears in the slow time-scale, suggesting that this
phenomena is the behavioral signature of a network that is able to transmit
not only large amounts of information between pairs of elements (high MIR) but
also overall within the network (high $H_{KS}$).

Note however, that since the evolving networks have a small number of
elements, the cost function cannot reach higher values and therefore, the
networks are not as optimal as they can be.  For that reason, we proceed now
to evolve larger networks, with $N$=32.

\begin{figure}[!h]
\centerline{\hbox{\psfig{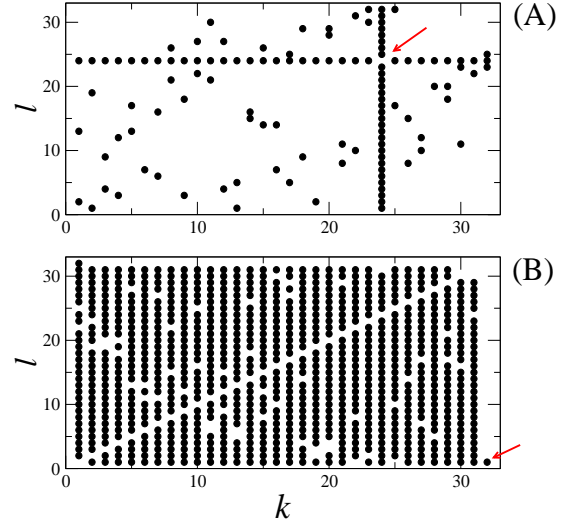}}}
\caption{{\footnotesize A point in this figure in the coordinate $k \times l$ means that 
the elements $S_k$ and $S_l$ are connected with equal couplings in a
bidirectional fashion. In (A), a 32 elements network, constructed by
maximizing the cost function ${\mathcal{B}}_1$ in Eq. (\ref{cond1}) and in
(B), 32 elements network, constructed by maximizing the cost function
${\mathcal{B}}_2$ in Eq. (\ref{cond2}). In (A), the network has the topology
of a perturbed star, a hub of neurons connected to all the other neurons,
where each outer neuron is sparsely connected to other neurons. The arrow
points to the hub. In (B),the network has the topology of a perturbed
all-to-all network, where elements are almost all-to-all connected. Note that
there is one element, the neuron $S_{32}$, which is only connected to one
neuron, the $S_{1}$. This isolated neuron is responsible to produce the large
spectral gap between the eigenvalues $\gamma_3$ and $\gamma_2$.  In (A), the
relevant eigenvalues are $\gamma_{31}$=4.97272,
$\gamma_{32}$=32, which produce a cost function equal to
${\mathcal{B}}$=5.43478. In (B), the relevant eigenvalues are
$\gamma_2$=0.99761, $\gamma_{3}$=27.09788, which produce a cost function equal
to ${\mathcal{B}}_2$=26.1628.}}
\label{optimal_geometries00}
\end{figure}

Maximization of the cost function ${\mathcal{B}}_1$ leads to the network
connectivity shown in Fig.
\ref{optimal_geometries00}(A) and maximization of 
the cost function ${\mathcal{B}}_2$ leads to the network connectivity shown in
Fig.
\ref{optimal_geometries00}(B). In (A), the network has the topology
of a perturbed star, a neuron connected to all the other outer neurons, thus a
hub, and each outer neuron is sparsely connected to other outer neurons. The
arrow points to the hub. In (B),the network has the topology of a perturbed
all-to-all network, where elements are almost all-to-all connected. Note that
there is one element, the neuron $S_{32}$, which is only connected to one
neuron, the $S_{1}$. This isolated neuron is responsible to produce the large
spectral gap between the eigenvalues $\gamma_3$ and $\gamma_2$.

\begin{figure}[!h]
\centerline{\hbox{\psfig{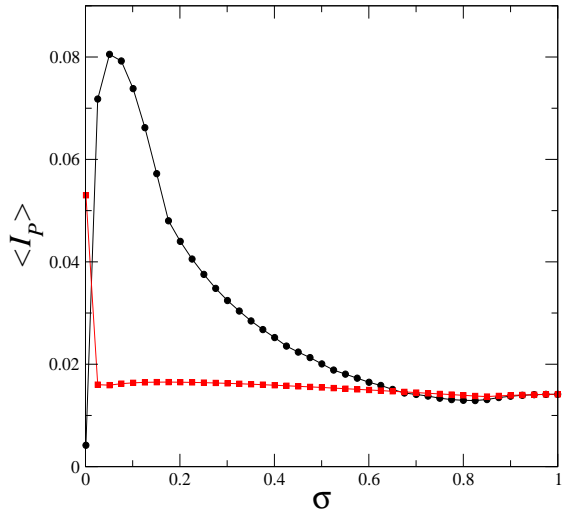}}}
\caption{$\langle I_P \rangle$ for the networks shown in
Fig. \ref{optimal_geometries00}(A-B) by circles and squares, respectively.}
\label{arbitrary_fig04}
\end{figure}

$\langle I_P \rangle$ for the network topology represented in
Fig. \ref{optimal_geometries00}(A) is shown in Fig. \ref{arbitrary_fig04} as
circles, and $\langle I_P \rangle$ for the network topology represented in
Fig. \ref{optimal_geometries00}(B) is shown in Fig. \ref{arbitrary_fig04} as
squares. We see that the star topology, whose connectivity is represented in
\ref{optimal_geometries00}(A), has larger $\langle I_P \rangle$ 
for a larger coupling strength than the topology
whose connectivity is represented in \ref{optimal_geometries00}(B).  Other
relevant parameters of the network whose topology is represented in
\ref{optimal_geometries00}(A) are $\sigma^2_{min}$=0.8468, 
$\sigma^3_{min}$=0.8249, , $\sigma^N_{min}$=0.0278, $\sigma_{CS}$=0.9762 and
for the topology represented in \ref{optimal_geometries00}(B) are
$\sigma^2_{min}$=0.8512, $\sigma^3_{min}$=0.042, $\sigma^N_{min}$=0.031, and
$\sigma_{CS}$=0.9761.

It is worth to comment that the neocortex is being simulated in the Blue Brain
project, by roughly creating a large network composed of many small networks
possessing the star topology. By doing that, one tries to recreate the way
minicolumnar structures \cite{malsburg1} are connected to minicolumnar
structures of the neocortex \cite{blue_gene}.  Each minicolumn can be
idealized as formed by a pyramidal neuron (the hub)
connected to its interneurons, the outer neurons in the star topology, 
which are responsible for the connections among this
minicolumn (small network) to others minicolumn. 
So, the used topology to simulate minicolumns is an optimal topology in what
concerns the transmission of information.

\subsection{Constructing a network by a given set of eigenvalues}
\label{pecora}

It is of general interest to assess if the eigenvalues obtained from the
method in Sec. \ref{metodo4} (in order to have a network Laplacian whose
eigenvalues maximize the cost function {$\mathcal{B}$}) can be used to
construct other networks (whose Laplacian preserve the eigenvalues)
maintaining still the properties here considered to be vital for information
transmission.

By a given set of eigenvalues, one can create a Laplacian matrix,
${\mathcal{G}^{\prime}}$, with non-zero real entries, using the method
described in Sec. \ref{metodo5}. The resulting network will preserve the
eigenvalues and the synchronous solution in Eq. (\ref{synchronous}), which
means that the values of $I_P^i$ of the topology created by the method in
Sec. \ref{metodo5} are equal to the values of the network topologies that
provide the set of eigenvalues, in the following example, the network
connectivities represented in Fig. \ref{optimal_geometries00}(A-B).

\begin{figure}[!h]
\centerline{\hbox{\psfig{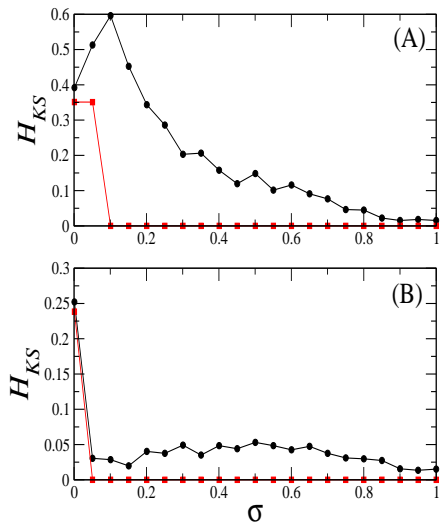}}}
\caption{$H_{KS}$ for a network evolved by the method in Sec.
\ref{metodo4}, in circles, and in squares, for a network  whose Laplacian is calculated
by the method in Sec. \ref{metodo5} in order for the Laplacian to generate the
same eigenvalues as the ones generated by the network Laplacian calculated by
the evolution technique. In (A), we consider the same network topology whose
connectivity is represented in Fig. \ref{optimal_geometries00}(A), and in (B),
we consider the same network topology whose connectivity is represented in
Fig. \ref{optimal_geometries00}(B). Note that in general, a Laplacian of an
active network whose elements are connected with different coupling strengths,
possess a smaller value of $H_{KS}$.}
\label{DD_fig7nova}
\end{figure}

In Fig. \ref{DD_fig7nova}(A-B), circles represent the values of $H_{KS}$ for
the network whose connectivities are represented in Figs.
\ref{optimal_geometries00}(A-B), and the squares represent this same quantity
for a network whose Laplacian is calculated by the method in
Sec. \ref{metodo5}, in order to preserve the same eigenvalues of the network
topologies represented in Fig. \ref{optimal_geometries00}(A-B).

The main difference between these two networks is that for the one constructed
by the method in Sec. \ref{metodo5}, when $\lambda^2$ becomes zero,
simultaneously $\lambda^1$ becomes also zero, a consequence of the fact that
all the neurons enter in a non-trivial but periodic oscillation.  In general,
however, both networks preserve the characteristics needed for optimal
information transmission: large amounts of MIR and $H_{KS}$, however, the ones
constructed by the evolution technique have larger $H_{KS}$, and possess a
larger MIR for larger ranges of the coupling strength. The network obtained by
the method in Sec. \ref{metodo5} is more synchronizable, a consequence of the
fact that the coupling strengths are non-equal \cite{stefano,jurgen}.

\section{Active networks formed by non-chaotic elements}\label{NC}

The purpose of the present work is to describe how information is transmitted
via an active media, a network formed by dynamical systems. There are three
possible asymptotic stable behaviors for an autonomous dynamical system:
chaotic, periodic, or quasi-periodic. A quasi-periodic behavior can be usually
replaced by either a chaotic or a periodic one, by an arbitrary
perturbation. For that reason, we neglect such a state and focus the attention
on active channels that are either chaotic or periodic.

The purpose of the present section is dedicated to analyze how a source of
information can be transmitted through active channels that are non-chaotic,
that is periodic, and that possess negative Lyapunov exponents.

Equation (\ref{capacity}) is defined for positive exponents. However, such an
equation can also be used to calculate an upper bound for the rate of mutual
information in systems that also possess negative Lyapunov exponents. Consider
first a one-dimensional contracting system being perturbed by a random
stimulus. Further consider that the stimulus changes the intrinsic dynamics of
this system. This mimics the process under which an active element adapts to
the presence of a stimulus.

Suppose the stimulus, $\theta_n$, can be described by a discrete binary random
source with equal probabilities of generating '0' or '1'.  Whenever $\theta_n
= 0$, the system presents the dynamics $x_{n+1}=x_n/2$, otherwise
$x_{n+1}=(1+x_n)/2$. It is easy to see that the only Lyapunov exponent of this
mapping, $\lambda_1$, which is equal to the conditional exponent, $\lambda^1$,
is negative. Negative exponents do not contribute to the production of
information. From Eq. (\ref{capacity}) one would arrive at $I_P$=0. However,
all the information about the stimulus is contained in the trajectory. If one
measures the trajectory $x_n$, one knows exactly what the stimulus was, either
a '0' or a '1'. The amount of information contained in the stimulus is
$\log{(2)}$ per iteration which equals the absolute value of the Lyapunov
exponent, $|\lambda_1|$.  In fact, it is easy to show that
$I_C=I_P=|\lambda^1|=|\lambda_1|=\log{(2)}$, or if we use the interpretation
of \cite{hayes}, $I_C=I_P=\lambda$, where $\lambda=|\lambda_1|$ is the
positive Lyapunov exponent of the time-inverse chaotic trajectory,
$x_{n+m},x_{n+m-1},\ldots,x_0$, which equals the rate of information
production of the random source. So, in this type of active communication
channel, one would consider in Eq. (\ref{capacity}) the positive Lyapunov
exponents of the time-inverse trajectory, or the absolute value for the
negative Lyapunov exponent.

Another example was given in \cite{baptista:2007}. In this reference we have
shown that a
chaotic stimulus perturbing an active system with a space contracting dynamics
(a negative Lyapunov exponent) might produce a fractal set. We assume that one
wants to obtain information about the stimulus by observing the fractal
set. The rate of information retrieved about the stimulus on this fractal set
equals the rate of information produced by the fractal set. This amount is
given by $D_1 |\lambda|$, where $D_1$ is the information dimension of the
fractal set and $|\lambda|$ the absolute value of the negative Lyapunov
exponent. In fact, $D_1 |\lambda|$ is also the rate of information produced by
the stimulus. So, if an active system has a space contracting dynamics, the
channel capacity equals the rate of information produced by the stimulus. In
other words, the amount of information that the system allows to be
transmitted equals the amount of information produced by the chaotic stimulus.

\section{The role of a time-dependent stimulus in an active network}\label{tds}

The most general way of modeling the action of an arbitrary stimulus
perturbing an active network is by stimulating it using uncorrelated white
noise. Let us assume that we have a large network with all the channels
operating in non-self-excitable fashion. We also assume that all the
transversal eigenmodes of oscillations except one are stable, and therefore do
not suffer the influence of the noise. Let us also assume that the noise is
acting only on one structurally stable (= far from bifurcation points)
element, $S_k$. To calculate the upper bound of the MIR between the element
$S_k$ and another element $S_l$ in the network, we assume that the action of
the noise does not alter the value of $\lambda^1$.  Then, the noise on the
element $S_k$ is propagated along the vibrational mode associated with the one
unstable transversal direction, whose conditional exponent is $\lambda^2$. As
a consequence, the action of the noise might only increase $\lambda^2$, while
not affecting the negativeness of all the other exponents ($\lambda^m$,
$m>2$), associated with stable transversal modes of oscillation.  That means
that the channels responsible for transmiting large amounts of information
(associated with $\lambda^m$, with $m$ large) will not be affected. So, for
such types of noises, Eq.  (\ref{capacity}) of the autonomous network is an
upper bound for the non-autonomous network.

Consider now a situation where the noise acts equally on all the elements of
an active network. The mapping of Eq. (\ref{mapa1}) was proposed as a way to
understand such a case. Consider the non-self-excitable map for $s$=-1. Note
that the term $\rho(x_n^2+y_n^2)$ that enters equally in all the maps has
statistical properties of an uniformly distributed random noise. Calculating
$I_P$ for $\rho=0$ (the noise-free map) we arrive at $I_P
\cong 2 \sigma$, for small $\sigma$, while the true MIR $I_C \cong 2(\sigma-\rho)$. These
results are confirmed by exact numerical calculation of the Lyapunov exponents
of Eq.  (\ref{mapa1}) as well as the calculation of the conditional exponents
of the variational equations.  So, this example suggests that Eq.
(\ref{capacity}) calculated for an autonomous non-perturbed network gives the
upper bound for the mutual information rate in a non-autonomous network.

\section{Discussions}

We have shown how to relate in an active network the rate of information that
can be transmitted from one point to another, regarded as mutual information
rate (MIR), the synchronization level among elements, and the connecting
topology of the network. By active network, we mean a network formed by
elements that have some intrinsic dynamics and can be described by classical
dynamical systems, such as chaotic oscillators, neurons, phase oscillators,
and so on.

Our main concern is to suggest how to construct an optimal network. A network
that simultaneously transmits information at a large rate, is robust under
couplings alterations, and further, it possesses a large number of independent
channels of communication, pathways along which information travels.

We have studied two relevant conditions that the eigenvalues of the Laplacian
matrix have to satisfy in order for one to have an optimal network. The
Laplacian matrix describes the coupling strengths among each element in the
active network. 

The two eigenvalues conditions are designed in order to produce networks that
are either self-excitable [maximizing Eq. (\ref{cond1})] or non-self-excitable
[maximizing Eq. (\ref{cond2})] (see definition of self-excitability in
Sec. \ref{metodo3}). Self-excitable networks have communication channels that
transmit information in a higher rate for a large range of the coupling
strength. Most of the oscillation modes in these networks are unstable, and
therefore, information is mainly propagated in a desynchronous environment.
Non-self-excitable networks have communication channels that transmit
information in a higher rate for a small range of the coupling strength,
however, they have channels that transmit reliable information in a moderate
rate for large range of coupling strengths.  Most of the oscillation modes in
these networks are stable, and therefore, information is mainly propagated in
a synchronous environment, a highly reliable environment for information
transmission.

Therefore, to determine the topology of an optimal network one does not need
to know information about the intrinsic dynamics of the elements forming the
network.

Once the network topology is obtained such that the eigenvalues of the
Laplacian matrix maximizes either the cost function in Eq. (\ref{cond1}) or
the one in Eq. (\ref{cond2}), the actual amount of information that can be
transmitted using the obtained topology will depend on the intrinsic dynamics
of the elements forming the network [$F$ in Eq. (\ref{element_dynamics})] and
also on the type of coupling [$H$ in Eq. (\ref{element_dynamics})], of only
two coupled elements [see Eq. (\ref{rescale})].

In the examples studied here, phase synchronization (PS) in the subspace
$(x,y)$ results in a great decrease of the KS-entropy (See Figs. \ref{DD_fig1}
and \ref{DD_fig2}) as well as of the MIR and $I_P$. However, a special type of
partial phase synchronization, the BPS, appears simultaneously when some
communication channel achieves its capacity. So, BPS \cite{baptista:2007} can
provide an ideal environment for information transmission, probably a
necessary requirement in the brain
\cite{lachaux,tass}.  Similarly, in networks of R\"ossler oscillators, a
type of non-self-excitable network, PS is the phenomenom responsible to
identify when the network is operating in a regime of high $MIR$
\cite{baptista:2005,baptista_CPL2006}. 

In order to construct an optimal network, we have used two approaches. One
based on a Monte Carlo evolving technique, which randomly mutates the network
topologies in order to maximize the cost functions in Eqs. (\ref{cond1}) and
(\ref{cond2}) (see Sec. \ref{metodo4}). We do not permit the existence of
degenerate eigenvalues. As a consequence $\gamma_N - \gamma_{N-1}$ as well as
$\gamma_3 - \gamma_{2}$ is never zero. The mutation is performed in order to
maximize the cost function, but we only consider network topologies for which
the value of the cost functions ${\mathcal{B}}_1$ and ${\mathcal{B}}_2$ remain
constant for about 10,000 iterations of the evolving technique, within
1,000,000 iterations. Even though more mutations could lead to networks that
have larger values of the cost function, we consider that a reasonably low
number of mutations would recreate what usually happens in real networks.  The
other approach creates an arbitrary Laplacian which reproduces a desired set
of eigenvalues.

Although both topologies provide larger amounts of MIR and $H_{KS}$, meaning
large network and channel capacities, the topology provided by the evolution
technique, which consider coupling strengths with equal strengths, is superior
in what concerns information transmission.  That agrees with the results of
Ref. \cite{baptista:2007} which say that networks composed by elements with
non-equal control parameters can transmit less information than networks
formed by equal elements, since networks whose coupling strengths are
non-equal can be considered to be a model for networks with non-equal control
parameters.

So, if brain-networks somehow grow in order to maximize the amount of
information transmission, simultaneously remaining very robust under coupling
alterations, the minimal topology that small neural networks must have should
be similar to the one in Fig. \ref{optimal_geometries00}(A), i.e., a network
with a star topology, presenting a central element, a hub, very well
connected to other outer elements, which are sparsely connected.

Even though most of the examples worked out here concern simulations performed
in a neural network of electrically coupled Hindmarsh-Rose neurons, our
theoretical approaches to find optimal topologies can be used to a large class
of dynamical systems, in particular also to networks of synaptically
(chemically) connected neurons. A neural network with neurons connected
chemically would also be optimal if one connect neurons by maximizing either
Eq. (\ref{cond1}) or Eq. (\ref{cond2}). The novelty introduced by the chemical
synapses is that it can enhance (as compared with the electrical synapses)
both the self-excitable (using excitable synapses) or the non-self-excitable
(using inibitory synapses) characteristic of the communication channels as
well as it can enhance $\langle I_P \rangle$
\cite{francois}. From the biological point-of-view, of course, the chemical
synapses provide the long-range coupling between the neurons. So, the
simulations performed here for the larger HR networks should be interpreted as
to simulations of a general active network, since neurons connected
electrically can only make nearest-neighbor connections.


\section{Methods}

\section{Calculating the MIR by symbolic encoding the trajectory}\label{metodo1}

The MIR between two neurons can be roughly estimated by symbolizing the
neurons trajectory and then measuring the mutual information 
from the Shannon entropy \cite{shannon} of the symbolic sequences.  From
\cite{shannon}, the mutual information between two signals $S_k$ and
$S_l$ is given by
\begin{equation}
I_{S}^{\prime}=H(S_k)-H(S_l|S_k).
\label{shannon1}
\end{equation}
$H(S_k)$ is the uncertainty about what $S_k$ has sent (entropy of the message),
and $H(S_l|S_k)$ is the uncertainty of what was sent, after observing
$S_l$.  In order to estimate the mutual information between two chaotic
neurons by the symbolic ways, we have to proceed with a non-trivial technique
to encode the trajectory, which constitutes a disadvantage of such technique
to chaotic systems.  We represent the time at which the $n$-th spike happens
in $S_k$ by $T_k^n$, and the time interval between the n-th and the (n+1)-th
spikes, by $\delta T_k^n$. A spike happens when $x_k$ becomes positive 
and we consider about 20000 spikes. We encode the spiking events using the
following rule. The $i$-th symbol of the encoding is a ``1'' if a spike is
found in the time interval $[i\Delta, (i+1) \Delta[$, and ``0'' otherwise.  We
choose $\Delta \in [\min{(\delta T_k^n)},
\max{(\delta T_k^n)}]$ in order to maximize $I_S^{\prime}$. Each neuron produces a
symbolic sequence that is split into small non-overlapping sequences of length
$L$=12. The Shannon entropy of the encoding symbolic sequence (in units of
bits), is estimated by $\max{H}|$ = -$\sum_m P_m \log_2 P_m$ where $P_m$
is the probability of finding one of the 2$^L$ possible symbolic sequences of
length $L$. The term $H(S_l|S_k)$ is calculated by
$H(S_l|S_k)$=$-H(S_l)+H(S_k;S_l)$, with $H(S_k;S_l)$ representing the Joint
Entropy between both symbolic sequences for $S_k$ and $S_l$.

Finally, the MIR (in units of bits/unit time), $I_S$, is calculated from 
\begin{equation}
I_S = \frac{I_S^{\prime}}{\Delta \times L}
\label{shannon}
\end{equation}

The calculation of the $I_S$ by means of Eq. (\ref{shannon}) should be
expected to underestimate the real value for the MIR. Since the HR neurons
have two time-scales, a large sequence of sequential zeros in the encoding
symbolic sequence should be expected to be found between two bursts of spikes
(large $\delta T_k^n$ values), which lead to a reduction in the value of
$H(S_k)$ followed by an increase in the value of $H(S_l|S_k)$, since there
will be a large sequence of zeros happening simultaneously in the encoding
sequence for the interspike times of $S_k$ and $S_l$.

\section{Self-excitability}\label{metodo3}

In Ref. \cite{baptista:2007} self-excitability was defined in the following
way.  An active network formed by $N$ elements, is said to be self-excitable
if $H_{KS}(N,\sigma) > H_{KS}(N,\sigma=0)$, which means that the KS-entropy of
the network increases as the coupling strength is increased. Thus, for non
self-excitable systems, an increase in the coupling strength among the
elements forming the network leads to a decrease in the KS-entropy of the
network.

Here, we adopt also a more flexible definition, in terms of the properties of
each communication channel. We define that a communication channel $c^i$ behaves
in a self-excitable fashion if $\lambda^i > \lambda^1$. It behaves in a
non-self-excitable fashion if $\lambda^i \leq \lambda^1$.

\section{Mutual Information Rate (MIR), channel
capacity, and network capacity}\label{metodo2}

In this work, the rate with which information is exchanged between two elements
of the network is calculated by different ways. Using the approaches of
Refs. \cite{baptista:2005,baptista:2007}, we can have an estimate of the
real value of the MIR, and we refer to this estimate as $I_C$. Whenever
we use Eq. (\ref{capacity}) to calculate the upper bound for the MIR, we will
refer to it as $I_P$. Finally, whenever we calculate the MIR through the
symbolic encoding of the trajectory as described in Sec. \ref{metodo1}, we refer to it
as $I_S$.

We define the {\it channel capacity} of a communication channel formed by two
oscillation modes depending on whether the channel behaves in a self-excitable
fashion or not. So, for the studied network, every communication channel
possess two channel capacities, the self-excitable capacity and the
non-self-excitable one. A channel $c^i$ operates with its self-excitable
capacity when $I_P^i$ is maximal, what happens at the parameter
$\sigma^{(i+1)*}$. It operates with its non-self-excitable capacity when
$\lambda^{i+1}=0$.

We also define the channel capacity in an average sense. In that case, the
averaged channel capacity is given by the maximal value of the average value
\begin{equation}
\langle I_P \rangle=\sum_{i=2}^{N}
\frac{1}{N-1}
|\lambda^1 - \lambda^{i}|,\label{averagedIP}
\end{equation}

The {\it network capacity} of a network composed of $N$ elements,
${\mathcal{C}}_N(N)$, is defined to be the maximum value of the
Kolmogorov-Sinai (KS) entropy, $H_{KS}$, of the network. For chaotic networks,
the KS-entropy, as shown by Pesin
\cite{pesin}, is the sum of all the positive Lyapunov
exponents. Notice that if $I$ denotes the MIR then 
\begin{equation}
I \leq H_{KS}
\label{condicaoI_H_KS}
\end{equation}

As shown in Ref. \cite{baptista:2007} and from the many examples treated here,
${\mathcal{C}}_N(N) \propto N$, and so, the network capacity grows linearly
with the number of elements in an active network.

\section{Understanding Eq. (\ref{capacity})}\label{network_maps}

Let us study Eq. (\ref{capacity}) using an analytical example. For an
introduction to the quantities shown here see Sec. \ref{metodo2}. Consider the
following two coupled maps:
\begin{eqnarray}
x_{n+1}&=&2x_n - \rho x_n^2 + 2s \sigma (y_n-x_n),\nonumber \\
y_{n+1}&=&2y_n - \rho y_n^2 + 2s \sigma (x_n-y_n),
\label{mapa1}
\end{eqnarray}
\noindent
with $\rho \geq 0$, $s=\pm 1$, and $x_n,y_n \in [0,1]$, which can be
accomplished by applying the $mod(1)$ operation.

\subsection{Positiveness of the MIR in Eq. (\ref{mapa1})}

Here, we assume that $\rho$=0. This map produces two Lyapunov exponents
$\lambda_1$=$\log{(2)}$ and $\lambda_2$=$\log{(2-4 s
\sigma)}$. Since this map is linear, the conditional exponents are equal to
the Lyapunov exponents.
 
Using the same ideas of Ref. \cite{baptista:2007}, actually an interpretation
of the way Shannon \cite{shannon} defines mutual information, the mutual
information rate, $I_P$, exchanged between the variables $x$ and $y$ is given
by the rate of information produced in the one-dimensional space of the
variable $x$, denoted as $H_x$, plus the rate of information produced in the
one-dimensional space of the variable $y$, denoted as $H_y$, minus the rate of
information production in the $(x,y)$ space, denoted as $H_{xy}$. But,
$H_x=H_y=\max{(\lambda_1,\lambda_2)}$, and
$H_{xy}$=$H_{KS}$=$\lambda_1+\lambda_2$, if $(\lambda_1,\lambda_2)>0$,
$H_{xy}$=$H_{KS}$=$\lambda_1$, if $\lambda_2<0$, and
$H_{xy}$=$H_{KS}$=$\lambda_2$, otherwise.

So, either $I_P=\lambda_1-\lambda_2$, case that happens for when $s=+1$, or
$I_P=\lambda_2-\lambda_1$, case that happens for when $s=-1$. If $s=+1$, the
larger the coupling strength, the smaller the KS-entropy, $H_{KS}$. If $s=-1$,
the larger the coupling strength, the larger $H_{KS}$. In fact, as we discuss
further, Eq. (\ref{mapa1}) for $s=-1$ is a model for a self-excitable channel,
and for $s=+1$ is a model for a non-self-excitable channel. In either case, the
MIR can be calculated by using the modulus operation as in
$I_P=|\lambda_1-\lambda_2|$. For larger networks, one can generalize such an
equation using the conditional exponents arriving to an equation of the form
as presented in Eq. (\ref{capacity}).

This equation points out to a surprising fact. Even when the level of
desynchronization in Eq. (\ref{mapa1}) is larger ($\lambda^2 >
\lambda^1$), which happens when $s=-1$, 
there is a positive amount of information being transferred between the two
variables.

\begin{figure}[!h]
\centerline{\hbox{\psfig{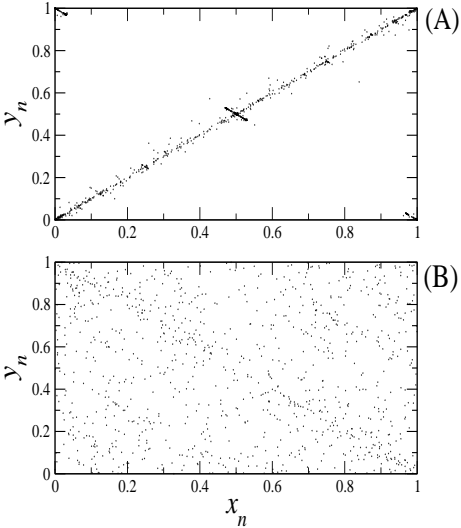}}}
\caption{Trajectory of Eq. (\ref{mapa1}) for $\sigma$=0.237 with 
$s=-1$ in (A) and $s=1$ in (B).}
\label{DD_fig6}
\end{figure}

In Figure \ref{DD_fig6}, we show the phase space of Eq. (\ref{mapa1}) for a
coupling strength equal to $\sigma$=0.237. In (A), we illustrate a typical
situation that happens in non-self-excitable channels ($s=+1$). The elements
become synchronous presenting a trajectory that most of the time lies on the
synchronization manifold defined by $x_n-y_n$=0. In (B), we show a typical
situation that happens in self-excitable channels ($s=-1$). The elements become
non-synchronous presenting a trajectory that lies on the transversal manifold
defined by $x_n+y_n-c$=0, with $c$ being a constant within the interval $c \in
[0,1]$.

In (A), by observing the variable $x_n$ one can correctly guess the
value of $y_n$  since $x_n \cong y_n $. Apparently, that is
not the case in (B): by observing the variable $x_n$, one might
have difficulty in guessing the value of the variable $y_n$, since $c
\in [0,1]$. Notice that the larger the amount of information being exchanged
between $x_n$ and $y_n$, the larger the chance that we guess correctly.  In
order to estimate the amount of information being exchanged between $x_n$ and
$y_n$, we proceed in the following way.

For the non-self-excitable channel ($s$=+1), we coarse-grain the phase space
in $L^2$ small squares. Each square has one side that represents an interval
of the domain of the variable $x_n$ and another side which is an interval of
the domain of the variable $y_n$. Calling $p_x^{(i)}$, the probability that a
trajectory point visits the interval $x_n=[(i-1)/L,i/L]$, with $i=1,\ldots,L$,
and $p_y^{(i)}$, the probability that a trajectory point visits the interval
$y_n=[(i-1)/L,i/L]$, and finally, $p_{x;y}^{(i,j)}$, the probability that a
trajectory point visits a square defined by $x_n=[(i-1)/L,i/L]$,
$y_n=[(j-1)/L,j/L]$, with $j=1,\ldots,L$, then, the MIR between $x_n$ and
$y_n$, denoted by $I$, is provided by {\small
\begin{equation}
I=-1/\log{(L)}[-\sum_i \log{(p_x^{(i)})} -\sum_i \log{(p_y^{(i)})} + \sum_{i,j}
\log{(p_{x;y}^{(i,j)})}].
\label{Ii}
\end{equation}
}\noindent
Notice that the evaluation of the MIR by Eq. (\ref{Ii}) underestimates the real
value for the MIR, since Eq. (\ref{mapa1}) is a dynamical system and the
information produced by the dynamical variables (for example the term $-\sum_i
\log{(p_x^{(i)})}$ that measures the information produced by the variable
$x_n$) should be provided by conditional probabilities, i.e., the probability
that a trajectory point has of visiting a given interval followed by another
interval, and so on, in fact the assumption used to derive
Eq. (\ref{capacity}).

\begin{figure}[!h]
\centerline{\hbox{\psfig{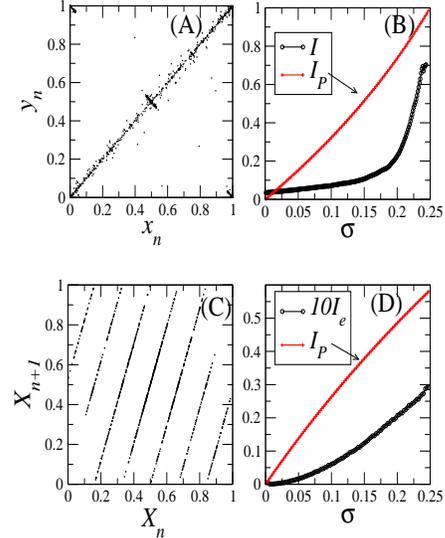}}}
\caption{Results for Eq. (\ref{mapa1}) with $\sigma0.237$ and 
$s$=-1 [shown in (A) and (B)] and for $s$=-1 [shown in (C) and (D)].  Phase
space of Eq. (\ref{mapa1}) and in (B), the MIR as calculated by
Eq. (\ref{capacity}) and as estimated by Eq. (\ref{Ii}). (C) Phase space of
Eq. (\ref{mapa1}) in the new coordinate frame $X_{n} vs. X_{n+1}$ and in (C),
the MIR as calculated by Eq. (\ref{capacity}) and as estimated by
Eq. (\ref{Ie}).}
\label{DD_fig5}
\end{figure}

In Fig. \ref{DD_fig5}(A), we show the phase space of Eq. (\ref{mapa1}) with
$s$=+1 and for $\sigma=0.237$. In Fig. \ref{DD_fig5}(B), we show by the plus
symbol, $I_P$, as calculated by Eq. (\ref{capacity}) and by circles, $I$, as
estimated by Eq. (\ref{Ii}).

For the self-excitable channel ($s$=-1) Eq. (\ref{Ii}) supplies a null 
MIR, and therefore, it can no longer be used. But, as discussed in
\cite{baptista:2007}, the MIR can be coordinate dependent, and one desires  
to have the coordinate that maximizes the MIR. Aiming at maximizing the MIR,
when the channel is of the self-excitable type, we transform Eq. (\ref{mapa1})
into an appropriate coordinate system, along the transversal manifold, where
most of the information about the trajectory position is located. We define
the new coordinate as $X_n = 1/2(x_n-y_n+1)$ and $X_{n+1} =
1/2(x_{n+1}-y_{n+1}+1)$.  The trajectory ($X_{n}, X_{n+1}$) in this new
coordinate system [for the same parameters as in Fig.
\ref{DD_fig6}(B)] is depicted in Fig. \ref{DD_fig5}(C).

The MIR being transferred between $X_{n}$ and $X_{n+1}$ is related to the
knowledge we acquire about $X_{n+1}$ by observing $X_n$, or vice-versa.
In Fig.
\ref{DD_fig5}(C), we can only be certain about the value of $X_{n+1}$,
when $X_n$ is close to either 0 or 1. 

To estimate the MIR, we recall that an encoded version of such a dynamical
system can be treated as a symmetric binary channel of communication. $X_n$ is
regarded as the transmiter and $X_{n+1}$ is regarded as the receiver. Whenever
the map in the transformed coordinates $X_{n} vs. X_{n+1}$ is non-invertible,
we consider that by making measures of the trajectory point $X_{n+1}$ one
cannot guarantee the exact position of the trajectory of $X_{n}$, which
constitutes an error in the transmission of information. Whenever the map is
invertible, by measuring the trajectory of $X_{n+1}$ one can surely know the
exact position of the trajectory $X_{n}$, which corresponds to a correct
transmission of information.  Calling, $p$ the probability at which the map is
invertible, then, the MIR between $X_n$ and $X_{n+1}$ is given by
\begin{equation}
I_e=1+(1-p)\log{(1-p)}+p\log{(p)}.
\label{Ie}
\end{equation}

The value of 10$I_e$ for Eq. (\ref{mapa1}) with $s=-1$ are shown in
Fig. \ref{DD_fig5}(D) by circles. The theoretical value, $I_P$, provided by
Eq. (\ref{capacity}) is shown by the plus symbol.

A final comment on the characteristics of a self-excitable channel and of a
non-self-excitable channel is that while in a self-excitable channel the
larger the synchronization level, the larger the MIR but the smaller the
KS-entropy, in a non-self-excitable channel the larger the desynchronization
level, the larger the MIR and the larger the KS-entropy. Note that
$H_{KS}$=$2\log{(2)}+\log{(1-2s \sigma)}$, for $\sigma<0.25$.

\subsection{Positiveness of the MIR for self-excitable channels
in the (non-linear) HR network}

To show that indeed $I_P^i$ should be positive in case of a self-excitable
channel in the HR network, one can imagine that in
Eq. (\ref{element_dynamics}) the coupling strength is arbitrarily small and
that $N$=2. At this situation, the Lyapunov exponent spectra obtained from
Eq. (\ref{variational}) are a first-order perturbative version of the
conditional exponents, and they appear organized by their strengths. One
arrives at $\lambda_1 \cong \lambda^2$ and $\lambda_2
\cong \lambda^1$, which means that the largest Lyapunov exponent equals the
transversal conditional exponent and the second largest Lyapunov exponent
equals the conditional exponent associated with the synchronous manifold,
i.e., the Lyapunov exponent of Eq. (\ref{synchronous}). Using similar
arguments to the ones in Refs.
\cite{baptista:2007,sara,baptista:2005}, we have that 
the MIR is given by the largest Lyapunov exponent minus the second largest,
and therefore, $I_C = \lambda_1-\lambda_2$, which can be put in terms of 
conditional exponents as 
$I_P \leq \lambda^2-\lambda^1$. 

\subsection{The inequality in Eq. (\ref{capacity})}

To explain the reason of the inequality in Eq. (\ref{capacity}), consider the
nonlinear term in Eq. (\ref{mapa1}) is non null and $s$=1, and proceeds as
further.

For two coupled systems, the MIR can be writen in terms of Lyapunov Exponents
\cite{baptista:2007,mendes}. For two coupled systems, 
the MIR can be exactly calculated by $I_C=\lambda_1-\lambda_2$, since
$\lambda^{\parallel} = \lambda_1$ and $\lambda^{\perp}= \lambda_2$, assuming
that both $\lambda_1$ and $\lambda_2$ are positive. Calculating the
conditional exponents numerically, we can show that $I_P \geq I_C$, and thus
$I_P$ is an upper bound for the MIR. For more details on this inequality, see
\cite{KS_entropy}

\section{Bust Phase Synchronization (BPS)}\label{metodo0}

Phase synchronization \cite{book_synchro} is a phenomenon defined by
\begin{equation}
|\Delta \phi(k,l)| = |\phi_k - m \phi_l| \leq r,
\label{phase_synchronization}
\end{equation}
\noindent
where $\phi_k$ and $\phi_l$ are the phases of two elements $S_k$ and $S_l$,
$m=\omega_l/\omega_k$ is a real number \cite{murilo_irrational}, where
$\omega_k$ and $\omega_l$ are the average frequencies of oscillation of the
elements $S_k$ and $S_l$, and $r$ is a finite, real number
\cite{baptista:2006}. In this work, we have used in
Eq. (\ref{phase_synchronization}) $m=1$, which means that we search for
$\omega_k:\omega_l$=1:1 (rational) phase synchronization
\cite{book_synchro}. If another type of $\omega_k:\omega_l$-PS is present, the
methods in Refs. \cite{baptista_PHYSICAD2005,tiago:2007,baptista:2006} can
detect it. 

The phase $\phi$ is a function constructed on a 2D subspace, whose trajectory
projection has proper rotation, i.e, it rotates around a well defined center
of rotation. So, the phase is a function of a subspace. Usually, a good 2D
subspace of the HR neurons is formed by the variables $x$ and $y$, and
whenever there is proper rotation in this subspace a phase can be calculated
as shown in Ref. \cite{tiago_PLA2007} by
\begin{equation}
\phi_s(t)=\int_0^t \frac{\dot{y}x-\dot{x}y}{{(x^2+y^2)}}dt.
\label{phase_xy}
\end{equation} 
If there is no proper rotation in the subspace $(x,y)$ one can still find
proper rotation in the velocity subspace $(\dot{x},\dot{y})$ and a
corresponding phase that measures the displacement of the tangent vector
\cite{baptista_PHYSICAD2005} can be calculated
as shown in Ref. \cite{tiago_PLA2007} by 
\begin{equation}
\phi_v(t)=\int_0^t \frac{\ddot{y}\dot{x}-\ddot{x}\dot{y}}{{(\dot{x}^2+\dot{y}^2)}}dt.
\label{phase_dxdy}
\end{equation} 
If a good 2D subspace can be found, one can also define a phase by means of
Hilbert transform, which basically transforms an oscillatory scalar signal
into a two component signal \cite{gabor}.  In the active network of
Eqs. (\ref{HR}) with an all-to-all topology and $N$=4, for the coupling
strength interval $\sigma \cong [0,0.05]$, the subspace $(x,y)$ has proper
rotation, and therefore, $\phi_s(t)$ is well defined and can be calculated by
Eq.  (\ref{phase_xy}). However, for this coupling interval, Eq.
(\ref{phase_synchronization}) is not satisfied, and therefore, there is no PS
between any pair of neurons in the subspace $(x,y)$.

For the coupling strength interval $\sigma \cong [0.05,0.24]$, the neurons
trajectories lose proper rotation both in the subspaces $(x,y)$ and
$(\dot{x},\dot{y})$.  In such a case, neither $\phi_s(t)$ nor 
$\phi_v(t)$ can be calculated. This is due to the fact
that the chaotic trajectory gets arbitrarily close to the neighborhood of the
equilibrium point $(x,y)$=$(0,0)$, a manifestation that a homoclinic orbit to
this point exists.

In fact, the Hilbert transform fails to provide the phase from either scalar
signals $x$ or $y$, since these signals do not present any longer an
oscillatory behavior close to the equilibrium point. In such cases, even the
traditional technique to detect PS by defining the phase as a function that
grows by 2$\pi$, whenever a trajectory component crosses a threshold cannot be
used. Since the trajectory comes arbitrarily close to the equilibrium point,
no threshold can be defined such that the phase difference between pairs of
neurons is bounded. Notice that by this definition the phase difference equals
$2\pi \Delta N$, where $\Delta N$ is the difference between the number of
times the trajectory of $S_k$ and $S_l$ cross the threshold. For the neural
networks, $\Delta N$ could represent the difference between the number of
spikes between two neurons. A spike is assumed to happen in $S_k$ if $x_k$
becomes positive.

In order to check if indeed PS exists in at least one subspace,
alternative methods of detection must be employed as proposed in Refs.
\cite{baptista_PHYSICAD2005,tiago:2007}.  In short, if PS exists in a 
subspace then by observing one neuron trajectory at the time the other bursts
or spikes (or any typical event), there exists at least one special curve,
$\Gamma$, in this subspace, for which the points obtained from these
conditional observations do not visit its neighborhood. A curve
$\Gamma$ is defined in the following way. Given a point $x_0$ in the attractor
projected onto the subspace of one neuron where the phase is defined, $\Gamma$
is the union of all points for which the phase, calculated from this initial
point $x_0$ reaches $n
\langle r
\rangle$, with
$n=1,2,3,\ldots,\infty$ and $\langle r \rangle$ a constant, usually
2$\pi$. Clearly an infinite number of curves $\Gamma$ can be
defined. For coupled systems with sufficiently close parameters that present
in some subspace proper rotation, if the points obtained from the conditional
observations do not visit the whole attractor projection on this subspace, one
can always find a curve $\Gamma$ that is far away from the conditional
observations. Therefore, for such cases, to state the existence of PS one just
has to check if the conditional observations are localized with respect to the
attractor projection on the subspace where the phase is calculated.

\begin{figure}
\centerline{\hbox{\psfig{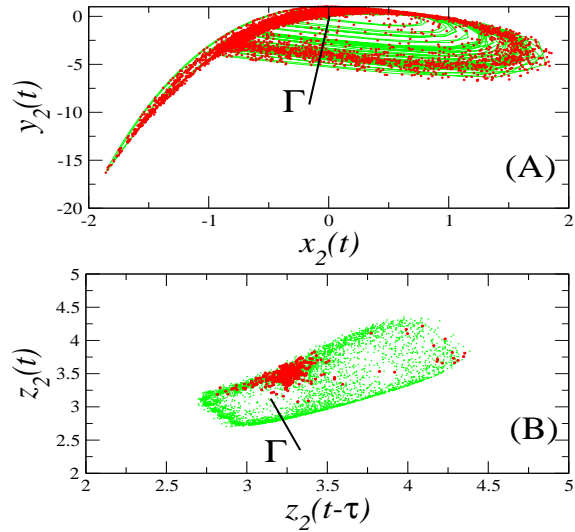}}}
\caption{{\footnotesize The network of Eqs. (\ref{HR}) with a star
configuration with $N$=4, and $\sigma$=0.265. The curve $\Gamma$, a continuous
curve transversal to the trajectory, is pictorially represented by the
straight line $\Gamma$.  (A) the green line represents the attractor
projection on the subspace $(x,y)$ of the neuron $S_2$, and red circles
represent the points obtained from the conditional observations of the neuron
$S_2$ whenever the neuron $S_4$ spikes. The point $(x,y)=(0.0)$ does
not belong to $\Gamma$. (B) Green dots represent the reconstructed attractor
$z_2(t) \times z_2(t - \tau)$, for $\tau$=30, and red circles represent the
points obtained from the conditional observation of neuron $S_2$,
whenever the reconstructed trajectory of the neuron $S_4$ crosses the
threshold line $z_4(t-\tau)=3.25$ and $z_4(t)>3$.}}
\label{fig_revised00}
\end{figure}

Conditional observations of the neuron trajectory $S_k$ in the subspace
($x,y$), whenever another neuron $S_l$ spikes, in the system modeled by Eqs.
(\ref{HR}) with a star coupling topology and $N$=4, are not localized with
respect to a curve $\Gamma$, for the coupling strength $\sigma <
\sigma_{PS}$. An example can be seen in Fig. \ref{fig_revised00}(A), for
$\sigma=0.265$. The set of points produced by the conditional observations are
represented by red circles, and the attractor by the green points. Therefore,
there is no PS in the subspace $(x,y)$. 

In order to know on which subspace PS occurs, we proceed in the following way.
We reconstruct the neuron attractors by means of the time-delay technique,
using the variable $z$. This variable describes the slow time-scale,
responsible for the occurrence of bursts.  The reconstructed attractor $z(t)
\times z(t - \tau)$ has proper rotation [see Fig. \ref{fig_revised00}(B)] 
and the points obtained from the conditional observations do not visit the
neighborhood of a curve $\Gamma$, then, there is PS in this subspace.  Indeed,
we find localized sets with respect to a curve $\Gamma$ in the reconstructed
subspace ($z(t)
\times z(t - \tau)$),  for $\sigma \geq 0.265$. So, $\sigma_{BPS}$=0.265.

So, for the coupling $\sigma = [\sigma_{BPS}, \sigma_{PS}[$, there is no PS in
the subspace $(x,y)$ but there is PS in the subspace of the variable $z$.  In
this type of synchronous behavior, the bursts are phase synchronized while the
spikes are not.  This behavior is regarded as bursting phase synchronization
(BPS).  For simplicity in the analyses, we say that BPS happens when for at
least one pair of neurons there is phase synchronization in the bursts. Phase
synchronization (PS) happens in the network when the average absolute phase
difference $$\frac{2}{N(N-1)} \sum_k \sum_l |\Delta \phi_L (k,l) |,$$ with
$k=1,N-1$ and $l=k+1,N$ among all the pairs of elements, is smaller than
$2\pi$, with the phases defined by either Eq. (\ref{phase_xy}) or
Eq. (\ref{phase_dxdy}), where the index $L$ represents either the index $s$ or
$v$. Further, we say complete synchronization (CS) takes place
\cite{pecora}, when the variables of one neuron equal the variables of all the
other neurons. 

For the analyses in this work, $\sigma_{BPS}$ represents the coupling
parameter for which BPS first appears, i.e., BPS exists if $\sigma \geq
\sigma_{BPS}$. $\sigma_{PS}$ represents the coupling parameter for which PS
first appears, i.e., PS exists if $\sigma \geq
\sigma_{PS}$. Finally, 
$\sigma_{CS}$ represents the coupling parameter for which CS first appears,
i.e., CS exists if $\sigma \geq
\sigma_{CS}$. There might exist particular parameters for 
which PS (or BPS) is lost even if $\sigma \geq \sigma_{PS}$ (resp. $\sigma
\geq \sigma_{BPS}$). But these parameters are not typical and we will ignore
them. For example, in the network composed by 6 elements with the
nearest-neighbor topology [Fig. \ref{DD_fig2}(B)], for $\sigma \cong 0.825$ PS
is lost.

Note that these phenomena happen in a hierarchical way organized by the
"intensity" of synchronization. The presence of a stronger type of
synchronization implies in the presence of other softer types of
synchronization in the following order: CS $\rightarrow$ PS $\rightarrow$ BPS.

\section{Evolutionary construction of a network}\label{metodo4}

In our simulations, we have evolved networks of equal bidirectional couplings
\cite{comment1}. That means that the Laplacian in Eq. (\ref{element_dynamics})
is a symmetric matrix of dimension $N$ with integer entries $\{0,1\}$ for the
off diagonal elements, and the diagonal elements equal to
$-\sum_j{\mathcal{G}}_{ij}$, with $i \neq j$.

Finding the network topologies which maximize ${\mathcal{B}}$ in
Eq. (\ref{cond1}) is impractical even for moderately large $N$. Figuring out
by "brute force" which Laplacian produces the desired eigenvalue spectra would
require the inspection of a number of $\frac{2^{N(N-1)/2}}{N!}$
configurations. To overcome this difficulty, Ref. \cite{evorene} proposed an
evolutionary procedure in order to reconstruct the network in order to
maximize some cost function.  Their procedure has two main steps regarded as
{\it mutation} and {\it selection}.  The mutation steps correspond to a random
modification of the pattern of connections.  The selection steps consist in
accepting or rejecting the mutated network, in accordance with the criterion
of maximization of the cost function ${\mathcal{B}}$, in Eq. (\ref{cond1}).

We consider a random initial network configuration, with $N$ elements, which
produce an initial Laplacian ${\mathcal{G}_0}$, whose eigenvalues produce a
value ${\mathcal{B}}_0$ for the cost function. We take at random one element
of this network and delete all links connected to it. In the following, we
choose randomly a new degree $k$ to this element and connect this element (in
a bidirectional way) to $k$ other elements randomly chosen. This procedure
generates a new network that possesses the Laplacian ${\mathcal{G}^{\prime}}$,
whose eigenvalues produce a value ${\mathcal{B}}^{\prime}$. To decide if this
mutation is accepted or not, we calculate $\Delta \epsilon
={\mathcal{B}}^{\prime} - {\mathcal{B}}_0$. If $\Delta \epsilon > 0$, the new
network whose Laplacian is ${\mathcal{G}^{\prime}}$ is accepted. If, on the
other hand, $\Delta
\epsilon < 0$, we still accept the new mutation, but with a probability $p(\Delta
\epsilon) = \exp(-\Delta/\epsilon T)$. If a mutation is accepted then the
network whose Laplacian is ${\mathcal{G}_0}$ is replaced by the network whose
Laplacian is ${\mathcal{G}^{\prime}}$.  

The parameter $T$ is a kind of ``temperature'' which controls the level of
noise responsible for the mutations. It controls whether the evolution process
converges or not. Usually, for high temperatures one expects the evolution
never to converge, since new mutations that maximizes ${\mathcal{B}}$ are
often not accepted. In our simulations, we have used $T
\cong 0.0005$.

These steps are applied iteratively up to the point when $|\Delta
\epsilon|=0$ for about 10,000 steps, being that we consider an evolution time
of the order of 1,000,000 steps. That means that the evolution process has
converged after the elapse of some time to an equilibrium state. If for more
than one network topology $|\Delta
\epsilon|=0$ for about 10,000 steps, we choose the network that has the larger
${\mathcal{B}}$ value.

This constraint avoids the task of finding the most optimal network
topology. However, we consider that a reasonably low number of mutations would
recreate what usually happens in real networks.

\section{Constructing a network from a set of eigenvalues}\label{metodo5}

Given a $N \times N$ Laplacian matrix ${\mathcal{G}}$, we can diagonalize it
by an orthogonal transformation, viz
\begin{equation}
 {\bf O}^T.{\mathcal{G}}.{\bf O} = {\bf \gamma} {\bf 1},  
\label{hussein0}
\end{equation}
\noindent
where ${\bf 1}$ represents the Unity matrix, ${\bf \gamma}$ represents the
vector that contains the set of eigenvalues $\gamma_i$ of ${\mathcal{G}}$
($i=1,\ldots,N$), and ${\bf O}$ is an orthogonal matrix, ${\bf O}.{\bf O}^T=
{\bf O}^T.{\bf O}={\bf 1}$, whose columns are constructed with the orthogonal
eigenvectors of ${\mathcal{G}}$, namely ${\bf
O}=[\vec{v}_1,\vec{v}_2,\ldots,\vec{v}_N]$. Accordingly,
\begin{equation}
{\mathcal{G}}={\bf O}.{\bf \gamma}{\bf 1}.{\bf O}^T, 
\label{hussein1}
\end{equation}
\noindent
which means that ${\mathcal{G}}$ can be decomposed into a multiplication of
orthogonal matrices. By using the spectral form of Eq. (\ref{hussein1}), the
Laplacian ${\mathcal{G}}$ can be calculated from
\begin{equation}
{\mathcal{G}}=\sum_{i=1}^N \vec{v}_i.\gamma_i.\vec{v}_i^T.
\label{hussein2}
\end{equation}

Any other Laplacian, ${\mathcal{G}^{\prime}}$, can be constructed by using the set of eigenvalues $\bf
\gamma$, viz
\begin{equation}
{\mathcal{G}^{\prime}}=\sum_{i=1}^N
\vec{v}^{\prime}_i.\gamma_i.\vec{v}^{\prime T}_i.
\label{pecora2}
\end{equation}
\noindent
Of course, in order for the active network that is constructed using
${\mathcal{G}^{\prime}}$ to present the synchronization manifold
$x_1=x_2=x_3=\ldots,=x_n$, the vector $\vec{v}^{\prime}_1$, with $N$ elements, is
given by $\vec{v}_1^{\prime T}$ = $\frac{1}{\sqrt{N}}[1,1,1,1,\ldots,1]$, and the other
vectors are found by choosing arbitrary vectors $\vec{v}^{\prime}_i$ which are made
orthogonal using the Gram-Schmidt technique.
 
\textbf{Acknowledgment} We thank C. Trallero who has promptly so many
times discussed with MSB related topics to this work. MSB thanks a stay
at the International Centre for Theoretical Physics (ICTP), where he had the
great opportunity to meet and discuss some of the ideas presented in this work
with H. Cerdeira and R. Ramaswamy. MSB also thanks K. Josi{\'c} for having
asked what would happen if the transversal conditional Lyapunovs were
larger than the one associated with the synchronization manifold and
T. Nishikawa for having asked what would happen if $s$ in Eq. (\ref{mapa1}) is
positive, two questions whose answer can be seen in Appendix
\ref{network_maps}. Finally, we would like to express our gratitude 
to L. Pecora, for clarifying the use of
the Spectal Theorem in the construction of a Laplacian matrix with a given set
of eigenvalues and for insisting in presenting a more rigorous argument
concerning the calculation of the conditional exponents.  This work is
supported in part by the CNPq and FAPESP. MSH is the Martin Gutzwiller Fellow
2007/2008.


\begin{thebibliography}{99}
\bibitem{diseases} Many pathological brain diseases, as Epilepsy, are
  associated with the appearance of synchronization.
\bibitem{comment1} Systems of bidirectional equal couplings can be
  considered as models of electrical gap junctions, a coupling that allows
  bidirectional flowing of information in neural networks.
\end{thebibliography}
\end{document}